\documentclass[aps,prd,preprint,groupedaddress]{revtex4-1}
\usepackage{graphicx}
\usepackage{amsmath}
\usepackage{amsfonts}
\usepackage{hyperref}
\usepackage{array}

\newcommand{\be}{\begin{equation}}
\newcommand{\ee}{\end{equation}}
\newcommand{\mpl}{M_\text{Pl}}
\newcommand{\half}{\frac{1}{2}}
\newcommand{\BP}[1]{\left[\Pi\right]^{#1}}
\newcommand{\TBP}[1]{\left[\Pi^{#1}\right]}
\newcommand{\pp}{\partial \pi}
\newcommand{\pb}{\bar{\pi}}
\newcommand{\MS}{M_\odot}

\begin{document}

\title{Galileon forces in the Solar System}

\author{Melinda Andrews}
\email[]{mgildner@sas.upenn.edu}
\author{Yi-Zen Chu}
\email[]{yizen.chu@gmail.com}
\author{Mark Trodden}
\email[]{trodden@physics.upenn.edu}

\affiliation{Center for Particle Cosmology, Department of Physics and Astronomy, University of Pennsylvania, 209 S 33rd Street, Philadelphia PA 19104-6396 USA}

\date{\today}

\begin{abstract}

We consider the challenging problem of obtaining an analytic understanding of realistic astrophysical dynamics in the presence of a Vainshtein screened fifth force arising from infrared modifications of General Relativity. In particular, we attempt to solve -- within the most general flat spacetime galileon model -- the scalar force law between well separated bodies located well within the Vainshtein radius of the Sun. To this end, we derive the exact static Green's function of the galileon wave equation linearized about the background field generated by the Sun, for the minimal cubic and maximally quartic galileon theories, and then introduce a method to compute the general leading order force law perturbatively away from these limits. We also show that the same nonlinearities which produce the Vainshtein screening effect present obstacles to an analytic calculation of the galileon forces between closely bound systems within the solar system, such as that of the Earth and Moon. Within the test mass approximation, we deduce that a large enough quartic galileon interaction would suppress the effect on planetary perihelion precession below the level detectable by even the next-generation experiments.

\end{abstract}

\maketitle

\newpage

%~~~~~~~~~~~~~~~~~~~~~~~~~~~~~~~~~~~~~~~~~~~~~~~~~~~~~~~~~~~~~~~~~~~~~~~~~~~~~~~~~~~~~~~~~~~~~~~~~~~~~~~~~~~~
\section{Introduction\label{s:intro}}
In exploring the possible space of allowed modifications to General Relativity (GR), one of the most stringent constraints is provided by precision tests within the solar system, which agree with GR to a high degree \cite{WillPPN}. If one is interested, for example, in infrared modifications of GR, in which late-time cosmic acceleration may be addressed, then one must satisfy these constraints while simultaneously seeking dynamics that depart strongly from GR on large scales. Thus, a viable modified gravity theory which explains cosmic acceleration is expected to display a {\it screening} mechanism that results in the behavior of the theory in high-density regions differing significantly from that on cosmological scales, where densities are relatively low. 

In this paper we wish to initiate the analytic study of long range forces mediated by galileons, a class of scalar particles appearing in various attempts to modify gravity at large distances and exhibiting what is known as the Vainshtein screening mechanism. (See \cite{nicolisgalileon} for the derivation of the action used here and \cite{Vainshtein1}-\cite{Vainshtein2} for discussion of the Vainshtein screening mechanism. Also see \cite{Hiramatsu:2012xj} for numerical work on the static Vainshtein-screened 2-body problem.) Possible astrophysical tests of galileon theories are discussed in \cite{tests1}-\cite{deRham:2012fg}, and the theory and its many extensions \cite{extensions1}-\cite{extensions19} have been of particular interest due to their cosmological consequences.

The primary hurdle to understanding both gravitational and Vainshtein screened forces is the presence of the nonlinear graviton and galileon self-interactions. In GR, gravitational forces within the solar system cannot be computed exactly but they can be treated perturbatively, because the Einstein-Hilbert Lagrangian $\frac{\mpl^2}{2} \sqrt{|g|} \mathcal{R}$ written in a weakly curved spacetime
\begin{align*}
g_{\mu\nu} \equiv \eta_{\mu\nu} + \frac{h_{\mu\nu}}{\mpl}, \qquad \mpl \equiv \frac{1}{\sqrt{8\pi G_N}},
\end{align*}
is a series taking the schematic form $\sum_{n=2}^\infty \partial^2 h^n/\mpl^{n-2}$. (That is, higher nonlinearities contain exactly two derivatives but higher powers of $h/\mpl$.) To lowest order, $h/\mpl$ scales as the typical Newtonian potential $G_N m/r \ll 1$ generated by the masses in the solar system, where $G_N$ is Newton's constant, $m$ is a typical mass and $r$ is the distance from the mass. Thus, each increasing order in nonlinearity scales as $h/\mpl \ll 1$, and is hence a small perturbation relative to the previous order. Moreover, the $1/$(distance) form of the Newtonian potential, and its nonlinear (and relativistic) corrections, is valid for any separation distance from a massive body within the solar system. In GR it is the Schwarzschild radius of the mass in question (when $h/\mpl \sim 1$) that characterizes when nonlinearities will dominate the dynamics. Even for the Sun, the most massive object in our solar system, its Schwarzschild radius is a mere 3 km while its physical radius is $7 \times 10^5$ km.

However, because the mass scale $\Lambda$ associated with the nonlinear self-interactions of galileons $\pi$ is much smaller than the mass scale $M_\text{pl} \sim 1/\sqrt{G_N}$ associated with gravitational interactions, the force law produced by galileons does not remain the same for all relevant separation distances. To see this, we first note that the galileon Lagrangian written in flat spacetime takes the schematic form 
\begin{equation*}
\sum_{n=2}^5 (\partial\pi)^2 (\partial^2\pi)^{n-2}/\Lambda^{3(n-2)}\ ,
\end{equation*}
where the degree of nonlinearity is measured by the powers of $\partial^2 \pi/\Lambda^3$ in a given term. Galileons couple to the trace of the stress-energy of matter, with an interaction Lagrangian $\pi T/\mpl$. For an isolated static body of mass $M$, we may define $1/\Lambda^3 \equiv (\mpl/M) r_v^3$, where $r_v$ is the Vainshtein radius of the object. Far away from $M$ the galileon potential is the familiar $\pi/\mpl \sim M/(\mpl^2 r)$ form. However, when one gets closer to the source $M$ than $r \lesssim r_v$ -- i.e. once $(\mpl/M) r_v^3 \partial^2 \pi \sim 1$ -- the nonlinearities will begin to dominate the dynamics. Typically, for galileons to be relevant cosmologically, $\Lambda \sim (\mpl H_0^2)^{1/3} \sim 1/10^3$ km ($H_0$ is the current Hubble parameter). This means the Sun's Vainshtein radius $r_v \sim 10^3$ light years and solar system dynamics takes place deep within the nonlinear regime of the Sun's (hypothetical) galileon field. Furthermore, the solar system is a multi-body system; to develop a quantitative understanding of its dynamics it is also necessary to  compute self-consistently the galileon forces exerted by the rest of the planetary bodies in the presence of the Sun. For instance, properly calculating the Earth-Moon galileon force is important if one wishes to use precision lunar laser ranging measurements to constrain the galileon's existence.

In this work, we approximate the Sun and its planetary companions as point masses and assume that the solar system is held together primarily by weak field gravity described by GR -- we will assume galileon forces are subdominant. We will focus only on the modification of solar system dynamics due to galileons, and thus attempt to solve the galileon theory in exactly flat spacetime.  This is justified as the interaction of galileons and gravity can be taken into account perturbatively, and we will show that these corrections are subdominant.  However, we do not take into account the effects of cosmological boundary conditions as in \cite{Babichev:2012re}.

 To capture the effects of nonlinearities we will first solve the galileon field $\bar{\pi}$ due to the Sun. We proceed to solve for the static Green's function of the galileon fluctuations about $\bar{\pi}$, and then use a field theoretic framework to examine the effective action of these point masses. This effective action framework is very similar in spirit to that developed in \cite{EFT} for the 2-body problem in GR (and extended in \cite{nbody} to the $n$-body case). We will exploit the fact that the solar system is a non-relativistic system and therefore, to zeroth order, the galileon potential between two point masses $M_{1,2}$ is simply $M_1 M_2/\mpl^2$ multiplied by the static Green's function of the galileon wave equation linearized about $\bar{\pi}$. Recently, in \cite{galileonwaves}, this static potential was solved exactly for the purely cubic galileon theory. Here, we will extend that work and find an exact solution for the maximally quartic galileon case. Using these exact solutions, we will also demonstrate how one may compute the general galileon force law using the perturbation theory approach developed in \cite{curvedG}.

This paper will be organized as follows: in Section~\ref{ss:background} we introduce the galileon theory and discuss the galileon force sourced by the Sun, including the resulting planetary perihelion precession. In Section~\ref{ss:propagators} we obtain the full static propagator for galileon interactions in the presence of a large central mass for both the minimum and maximum quartic interaction strengths allowed by stability requirements, and then extend these results to more generic parameter values using perturbation theory.  In Section~\ref{ss:diagrams} we explain the power counting of Feynman diagrams in the field theoretic framework and present the resulting form for the effective action as an expansion in two small parameters.  In Section~\ref{ss:outsiderv} we provide concrete results in the region outside the Vainshtein radius of the large central source, followed in Section~\ref{ss:breakdown} by a discussion of why this calculation should not be trusted in the region inside the Vainshtein radius.  We discuss the results in Section~\ref{s:discussion}.

%~~~~~~~~~~~~~~~~~~~~~~~~~~~~~~~~~~~~~~~~~~~~~~~~~~~~~~~~~~~~~~~~~~~~~~~~~~~~~~~~~~~~~~~~~~~~~~~~~~~~~~~~~~~~
%~~~~~~~~~~~~~~~~~~~~~~~~~~~~~~~~~~~~~~~~~~~~~~~~~~~~~~~~~~~~~~~~~~~~~~~~~~~~~~~~~~~~~~~~~~~~~~~~~~~~~~~~~~~~
\section{Force Laws\label{s:forcelaws}}

%~~~~~~~~~~~~~~~~~~~~~~~~~~~~~~~~~~~~~~~~~~~~~~~~~~~~~~~~~~~~~~~~~~~~~~~~~~~~~~~~~~~~~~~~~~~~~~~~~~~~~~~~~~~~
\subsection{The Background\label{ss:background}} 

%~~~~~~~~~~~~~~~~~~~~~~~~~~~~~~~~~~~~~~~~~~~~~~~~~~~~~~~~~~~
\subsubsection{Theory under consideration\label{sss:theory}} 

We wish to examine the following theory to understand how Vainshtein screened scalar forces impact the dynamics of astrophysical systems:
\begin{equation}
\label{TotalAction}
S \equiv S_\pi + S_\text{point particles}\, .
\end{equation}
As derived in \cite{nicolisgalileon}, $S_\pi$ encodes the dynamics of a scalar field $\pi$ with derivative self-interactions consistent with a galilean shift symmetry $\pi \to \pi + b_\mu x^\mu + c$, and yields second-order equations of motion. The action is given by
\begin{equation}
S_\pi = \int d^4 x \left(\sum_{i=2}^5 \frac{\alpha_i}{\Lambda^{3(i-2)}} \mathcal{L}_i + \frac{\pi}{\mpl} T\right) \, ,
\label{eq:quinticaction}
\end{equation}
where, defining the matrix $\Pi^\mu_{\phantom{\mu}\nu} \equiv \partial^\mu \partial_\nu \pi$ and the notation $[A] \equiv A^\mu_{\phantom{\mu}\mu}$ (for any matrix $A$),
\begin{align}
\mathcal{L}_2 &= -\half \pp \cdot \pp \\
\mathcal{L}_3 &= -\half [\Pi] \pp \cdot \pp \\
\mathcal{L}_4 &= -\frac{1}{4} \left( \BP{2} \pp \cdot \pp - 2 [\Pi] \pp \cdot \Pi \cdot \pp - \TBP{2} \pp \cdot \pp + 2\, \pp \cdot \Pi^2 \cdot \pp \right) \\
\mathcal{L}_5 &= -\frac{1}{5} \Big( 
\BP{3} \pp \cdot \pp - 3 \BP{2} \pp \cdot \Pi \cdot \pp - 3 [\Pi] \TBP{2} \pp \cdot \pp + 6 [\Pi] \pp \cdot \Pi^2 \cdot \pp \nonumber\\
&\qquad\qquad 
+ 2 \TBP{3} \pp \cdot \pp + 3 \TBP{2} \pp \cdot \Pi \cdot \pp - 6\, \pp \cdot \Pi^3 \cdot \pp
\Big)\, .
\end{align}
We have chosen the coupling to matter $\left(\pi/\mpl\right) T$, which arises naturally from DGP-like models and is the lowest-order coupling term.  This choice is galilean-invariant in flat spacetime and for an external source; away from these assumptions the galilean symmetry is broken but the theory is nevertheless interesting as the simplest realization of the Vainshtein screening mechanism. Another possible coupling which arises in the decoupling limit of massive gravity is the disformal coupling $\left(\partial_\mu \pi \partial_\nu \pi/\mpl^4\right) T^{\mu\nu}$.  This coupling is parametrically smaller than the coupling to the trace of the stress-energy tensor so long as the galileon is sub-Planckian and quantum corrections to the galileon terms are irrelevant ($\alpha_q = \partial^2 / \Lambda^2 \ll 1$ \cite{extensions7}):
\begin{equation}
\frac{\partial_\mu \pi \partial_\nu \pi}{\mpl^4} T^{\mu\nu} \sim \frac{\pi}{\mpl} T \left( \alpha_q \left(\frac{\Lambda}{\mpl}\right)^2 \frac{\pi}{\mpl} \right)
\end{equation}
The disformal coupling is not only higher order, it is also zero in the case of a static, nonrelativistic source such as we consider here.  As discussed in \cite{tests2}, however, this coupling is important for lensing calculations (for which the lowest-order coupling is zero).

We consider as matter a collection of point particles whose motion is left arbitrary, aside from a large central mass $M_\odot$ which is pinned at the origin:
\begin{equation}
S_\text{point particles} = -M_\odot \int dt_M -\sum_{a=1}^N m_a \int dt_a \sqrt{-\eta_{\mu\nu} v_a^\mu v_a^\nu}\, .
\label{eq:matter_action}
\end{equation}
Via the standard definition $T_{\mu\nu} = -\frac{2}{\sqrt{-\eta}} \frac{\delta S_\text{pp}}{\delta \eta^{\mu\nu}}$ the trace of the stress-energy tensor to which the galileon couples is
\begin{equation}
T = -M_\odot \delta^3(\vec{x}) - \sum_{a=1}^N m_a \delta^3(\vec{x}-\vec{x}_a(t)) \sqrt{1-v_a(t)^2} \, .
\label{eq:T}
\end{equation}

{\it Strategy} \qquad Our approach to understanding galileon forces between well separated bodies lying deep within the Vainshtein radius of the central mass $\MS$ is as follows. We shall first solve for the galileon profile $\pb$ generated by $\MS$. This means setting to zero all the $\{m_a\}$ and solving the resulting $\pi$-equation from the variation of \eqref{TotalAction} (note that the shift symmetry ensures that the equation of motion is a total derivative, and thus can be integrated to obtain an algebraic equation for $\bar{\pi}'$ \cite{nicolisgalileon})
\begin{equation}
\label{BackgroundEquation}
\alpha_2 \left(\frac{\bar{\pi}'}{r}\right) + 2 \frac{\alpha_3}{\Lambda^3} \left(\frac{\bar{\pi}'}{r}\right)^2 + 2 \frac{\alpha_4}{\Lambda^6} \left(\frac{\bar{\pi}'}{r}\right)^3 = \frac{M_\odot}{4\pi \mpl r^3}\, ,
\end{equation}
which we rewrite as 
\begin{equation}
y + 2y^2 + 2x y^3 = \frac{1}{8 z^3}
\end{equation}
in terms of $y\equiv \frac{\alpha_3}{\alpha_2 \Lambda^3} \left( \frac{\bar{\pi}'}{r} \right)$, $x \equiv \frac{\alpha_2 \alpha_4}{\alpha_3^2}$, and $z^3 \equiv \frac{\pi \alpha_2^2}{2 \alpha_3} \left(\frac{r}{r_v} \right)^3$.
The cubic equation for $y$ has only one solution that is both real and satisfies the boundary condition that $y(z\to0) = 0$:
\begin{equation}
y = 2\frac{\sqrt{1-\frac{3}{2} x}}{3x} \left[{\cos \brace \cosh} \left(\frac{\theta (z)}{3}\right) - {\cos \brace \cosh}\left(\frac{\theta(0)}{3}\right) \right] \ ,
\end{equation}
where
\begin{equation}
\theta(z) = {\cos^{-1} \brace \cosh^{-1}}\left(\frac{-32+72 x+27 x^2 z^{-3}}{32 \left(1-\frac{3}{2} x\right)^{3/2}}  \right)
\end{equation}
and $\cos$ or $\cosh$ is chosen such that $\theta(z)$ is real.

We then carry out perturbation theory about $\pb$, replacing in \eqref{TotalAction}
\begin{equation}
\pi(t,\vec{x}) = \bar{\pi}(r) + \phi(t,\vec{x}) \, .
\label{phiexpansion}
\end{equation}
Isolating the quadratic-in-$\phi$ terms of \eqref{TotalAction}, we obtain the general form
\begin{equation}
\label{KineticTerms}
S_\text{kin} \equiv -\half \int d^4x \sqrt{-\eta} \left[ -K_t(r) (\partial_t\phi)^2 + K_r(r) (\partial_r\phi)^2 + K_\Omega(r) (\partial_\Omega\phi)^2 \right] \nonumber \, ,
\end{equation}
where $K_t(r)$, $K_r(r)$ and $K_{\Omega}(r)$ are functions of the background field $\bar{\pi}(r)$ given by (with occurrences of $\partial_r y$ eliminated using the equation of motion)
\begin{align}
K_t(r) &= \alpha_2 \left( \frac{ 1 + 4y + 12(1-x)y^2 + 24\left( x-2\frac{\alpha_2^2 \alpha_5}{\alpha_3^3} \right)y^3 + 12\left( 3x^2 - 4\frac{\alpha_2^2 \alpha_5}{\alpha_3^3} \right)y^4 }{ 1 + 4y + 6xy^2 } \right)    \nonumber \\
K_r(r) &= \alpha_2 \left( 1 + 4y + 6xy^2 \right)   \\
K_\Omega(r) &= \alpha_2 \left( \frac{ 1 + 2y + 2(2-3x)y^2 }{ 1 + 4y + 6xy^2 } \right) \, . \nonumber 
\end{align}
For an arbitrary static, spherically symmetric background, we may re-express the action as the kinetic term of a massless scalar field in a curved spacetime (see \cite{Babichev:2007dw} for a prior example of this procedure), namely
\begin{equation}
S_\text{kinetic} = -\half \int d^4 x \sqrt{-\tilde{g}} \tilde{g}^{\mu\nu} \partial_\mu \phi \partial_\nu \phi \, ,
\label{eq:S_kin_general}
\end{equation}
with the effective metric 
\begin{equation}
\tilde{g}_{\mu\nu} \equiv \text{diag} \left( - \sqrt{\frac{K_r}{K_t}} K_\Omega, \sqrt{\frac{K_t}{K_r}} K_\Omega, r^2 \sqrt{K_t K_r}, r^2 \sin^2\theta \sqrt{K_t K_r} \right) \, .
\label{eq:g_eff_general}
\end{equation}
The first order of business is then to solve the static Green's function of the resulting massless scalar wave operator,
\begin{equation}
\widetilde{\Box}_x G\left( \vec{x},\vec{y} \right)
= \widetilde{\Box}_y G\left( \vec{x},\vec{y} \right) 
= - \frac{\delta^{(3)}(\vec{x}-\vec{y})}{\sqrt[4]{-\tilde{g}(x)} \sqrt[4]{-\tilde{g}(y)}} \, .
\label{eq:Green's_fn}
\end{equation}
where $\widetilde{\Box} \equiv \widetilde{g}^{\mu\nu} \widetilde{\nabla}_\mu \widetilde{\nabla}_\nu$ and $\widetilde{\nabla}$ is the covariant derivative with respect to the effective metric in \eqref{eq:g_eff_general}. Strictly speaking, to probe the dynamical content of our scalar equation linearized about $\pb$, one would need to solve the full time dependent retarded Green's function $G_\text{ret}(x,y)$. However, this is not an easy task. Therefore, because we are interested primarily in the forces between planetary bodies moving at speeds much less than unity (relative to the solar system's center of energy), we shall in this paper seek its static limit, 
\begin{equation*}
G(\vec{x},\vec{y}) \equiv \int_{-\infty}^\infty d x^0 G_\text{ret}(x,y) = \int_{-\infty}^\infty d y^0 G_\text{ret}(x,y) \ .
\end{equation*}

Before proceeding, let us observe from \eqref{BackgroundEquation} that the quintic galileon self-interactions do not contribute to the background field $\pb$. In fact, as discussed in \cite{nicolisgalileon}, any time-independent galileon $\pi$ trivially satisfies the portion of the equations arising from $\mathcal{L}_5$. This (as expressed by the $\alpha_5$-independence of $K_r$ and $K_\Omega$) means the static Green's function obtained for $\alpha_4 = 0$ in \cite{galileonwaves} and the maximally quartic case we shall consider in Section~\ref{sss:maximalquartic} are also solutions in the presence of an arbitrary non-zero $\alpha_5$.  The quintic interactions become relevant, however, for the full time-dependent Green's function.

%~~~~~~~~~~~~~~~~~~~~~~~~~~~~~~~~~~~~~~~~~~~~~~~~~~~~~~~~~~~
\subsubsection{Precession of Mercury\label{sss:precession}} 

We begin by asking: what is the contribution to the precession of perihelia of planetary orbits due to the galileon field of the Sun? This has been calculated in \cite{cubicprecession} for galileons with only the cubic interaction term; here we extend the calculation to the case of the full galileon theory with cubic, quartic, and quintic interactions. To zeroth order, the galileon field in the solar system is primarily governed by the field due to the Sun, \eqref{BackgroundEquation}. As the orbits of the planets are well within the Sun's Vainshtein radius, the most nonlinear term dominates and hence the background solution is (up to an additive constant) well approximated by
\begin{equation}
\bar{\pi} = \frac{M_\odot r}{2(\pi \alpha_4)^{1/3} \mpl r_v^2} \, .
\end{equation}
Thus, we wish to consider the perihelion precession induced by a potential 
\begin{align}
\Psi = -\frac{G M_\odot}{r} + \frac{\bar{\pi}}{\mpl} \sim G M_\odot \left( -\frac{1}{r} + C r \right), \qquad C = \frac{8\pi}{2(\pi \alpha_4)^{1/3} r_v^2} \, .
\end{align}
Following the standard procedure for such calculations as found in any GR textbook, we find that to second order in the eccentricity $e$ for a planet with semi-major axis $a$, the perihelion precession induced by the galileons relative to that induced by GR is
\begin{equation}
\frac{\Delta\phi_\text{galileon}}{\Delta\phi_\text{GR}} = - \frac{8 \pi^{2/3}}{3\alpha_4^{1/3}}\frac{a^3}{r_s r_v^2} \ ,
\end{equation}
with $r_s$ the Schwarzschild radius of the Sun.  The galileon-induced precession thus decreases the precession per orbit from GR by an amount suppressed by the distance ratio $a^3/(r_s r_v^2)$.  For the Sun-Mercury system, this is an effect at $\mathcal{O}(10^{-10}) \Delta\phi_{GR}$, and hence is unobservable by all current and near-future methods of measurement. Note that this is the correct result only if the quartic interaction term dominates the cubic one at the orbit of the planet in question, or in other words, when $r_\text{Mercury} \ll r_{34}$ (see \eqref{VainshteinRadii} below). If we wish to consider the crossover from this behavior to the cubic-dominated precession which is several orders of magnitude larger, we must consider the full galileon force law.

This precession calculation treats every object in the solar system, other than the Sun itself, as a test body with negligible mass. However, due to the highly nonlinear nature of the galileon field in the solar system, it is important to go beyond this one body problem. We proceed henceforth to investigate the dynamics of objects due to galileon interactions between such objects within the background field of a large central mass.

%~~~~~~~~~~~~~~~~~~~~~~~~~~~~~~~~~~~~~~~~~~~~~~~~~~~~~~~~~~~~~~~~~~~~~~~~~~~~~~~~~~~~~~~~~~~~~~~~~~~~~~~~~~~~
\subsection{Forces Between Smaller Masses\label{ss:propagators}} 

To lowest order in galileon self-interactions, the scalar force between two test masses in the presence of the Sun can be determined from the curved-space Green's function with the effective metric as obtained from (\ref{KineticTerms}) and (\ref{eq:g_eff_general}).
We seek a full Green's function solution as an expansion in spherical harmonics via the ansatz
\begin{equation}
G(\vec{r},\vec{r}')=\sum_{\ell=0}^\infty \sum_{m=-\ell}^\ell R_{>,\ell}(r_>) R_{<,\ell}(r_<) Y_\ell^m (\Omega) Y_\ell^m (\Omega')^* \, ,
\end{equation}
where $r_>$ and $r_<$ are, respectively, the larger and smaller of $r$ and $r'$. Henceforth, $R_\ell(r)$ should be understood to mean the piecewise-defined function that is $R_{>,\ell}$ when $r=r_>$ and $R_{<,\ell}$ when $r=r_<$.  The main obstacle to finding a solution is then whether or not the radial equation
\begin{equation}
\left[K_r(r) R_\ell''(r) + K_r(r) \left(2 + \frac{r K_r'(r)}{K_r(r)}\right) \frac{1}{r} R_\ell'(r) - K_\Omega(r) \frac{\ell(\ell+1)}{r^2} R_\ell(r) \right] R_\ell(r') = -\frac{1}{r^2} \delta(r-r')
\label{eq:Rldiffeq}
\end{equation}
is solvable. Because $\bar{\pi}''$ in the $K_i$s can be exchanged for $y = \bar{\pi}'/r$ via the equation of motion, (\ref{eq:Rldiffeq}) depends on the background field solely through $y$, for arbitrary choices of the parameters $\alpha_i$. 

First note that the stability requirement that all the $K_i$s be positive restricts the viable parameter space to
\begin{equation}
\hfill \alpha_2 > 0 \hspace{1cm}
\alpha_3 \ge \sqrt{\frac{3}{2} \alpha_2 \alpha_4} \hspace{1cm}
\alpha_4 \ge 0 \hspace{1cm}
\alpha_5 \le \frac{3\alpha_4^2}{4\alpha_3} \ ,\hspace{1cm}
\end{equation}
as discussed in \cite{nicolisgalileon} (this analysis is modified for a spacetime which is not asymptotically Minkowski - see \cite{Babichev:2012re}). We will focus on the case where $\alpha_5=0$, since the quintic term has little effect on the analysis - it does not affect the background solution and only appears in $K_t$ and hence is irrelevant for the static Green's function. 

Note also that there are three different distance regimes relevant to the problem. The first is the regime far from the Sun (i.e, the central mass) beyond which the nonlinearities are unimportant, the second is the intermediate-distance regime where the cubic term dominates the dynamics, and the final one is the near-source regime where the quartic term dominates over the cubic term in determining the dynamics. The linear-nonlinear transition happens at a scale $r_{23}$ and the cubic-quartic transition at a second scale $r_{34}$, as defined by:
\begin{equation}
\label{VainshteinRadii}
r_{23}^3 = \frac{\alpha_3 M}{\alpha_2^2 \mpl \Lambda^3}\,,
\hspace{3cm}
r_{34}^3 = x^2 r_{23}^3\,,
\end{equation}
where it is convenient to define a new parameter $x$ controlling the relative importances of the cubic and quartic terms, via
\begin{equation}
0 \le x \equiv \dfrac{\alpha_2 \alpha_4}{\alpha_3^2} \le \frac{2}{3} \, .
\end{equation}
The bounds on the magnitude of $x$ are a consequence of the stability conditions above.

\begin{figure}[ht]
\includegraphics[width=\textwidth]{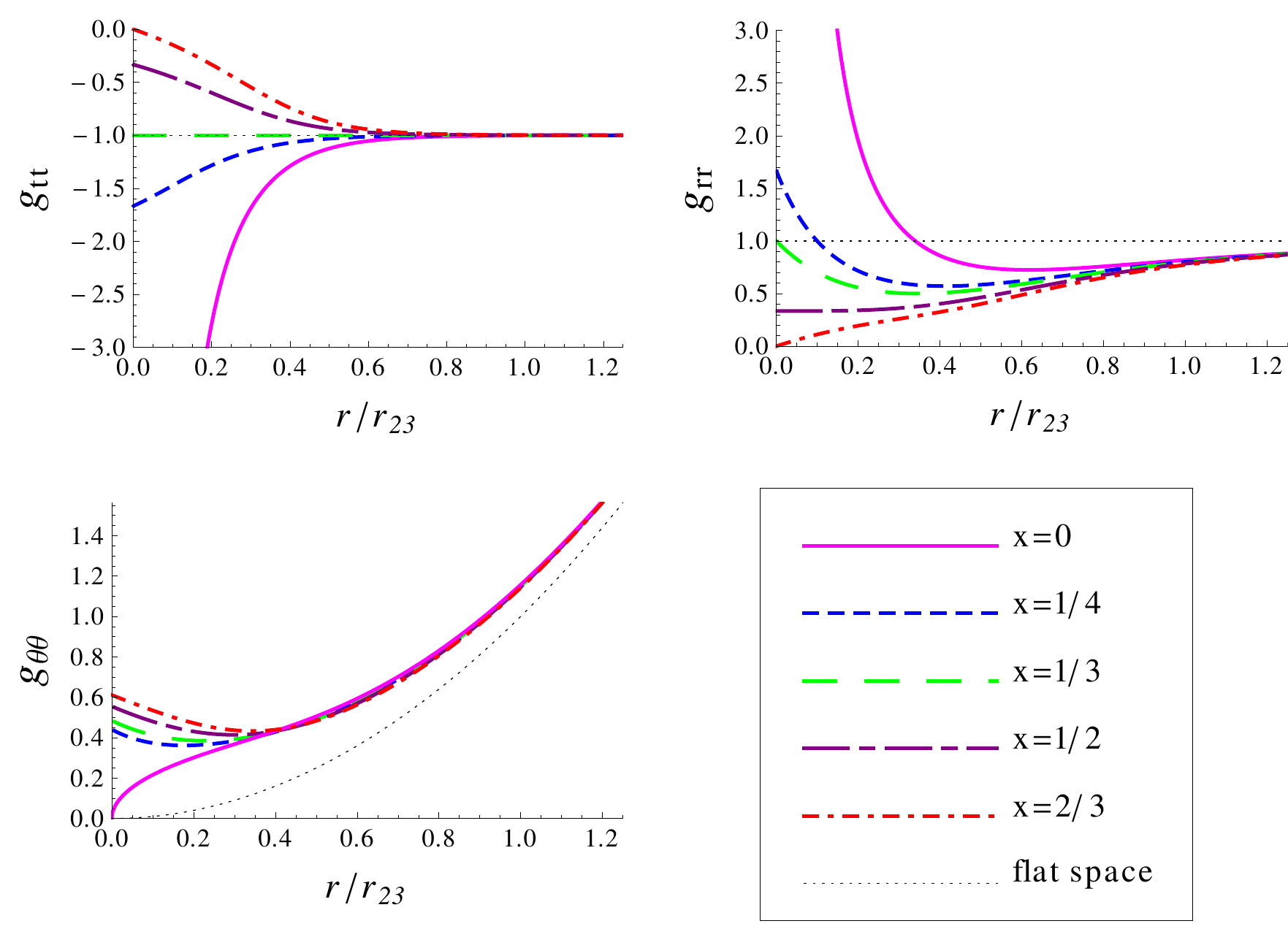}
\caption{\label{fig:metric} The effective metric seen by galileon fluctuations about the spherically-symmetric background for various relative strengths $x \equiv \alpha_2 \alpha_4/\alpha_3^2$ of the cubic and quartic terms.}
\end{figure}

\begin{figure}[ht]
\includegraphics[width=\textwidth]{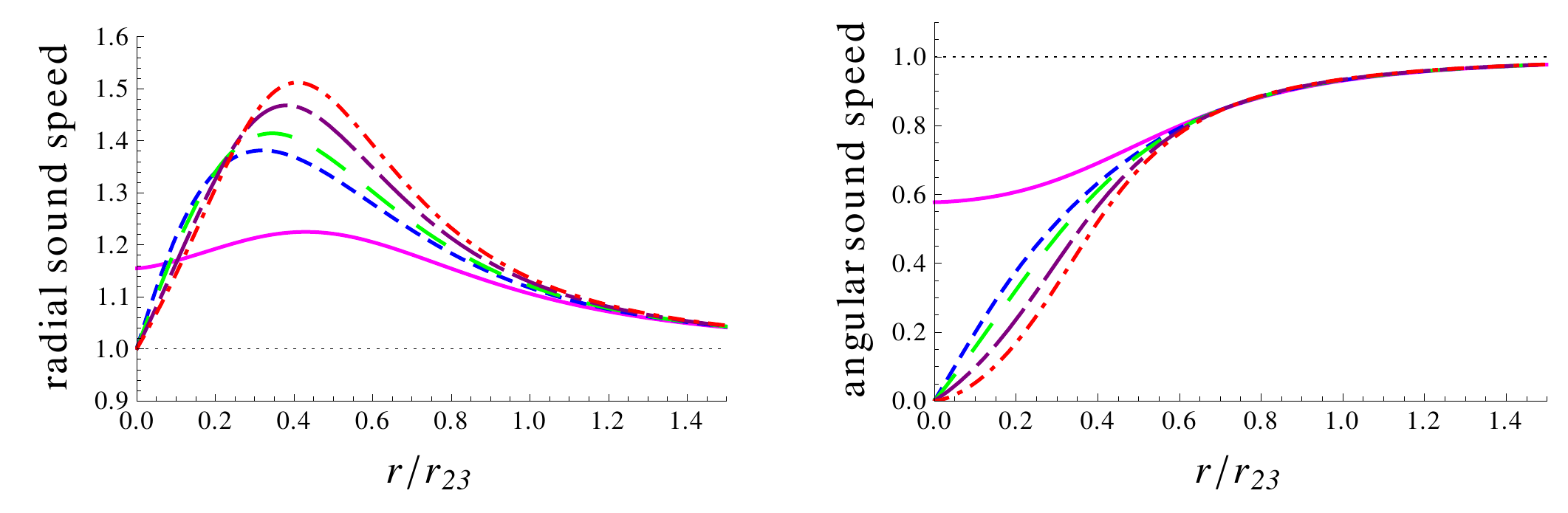}
\caption{\label{fig:speed} The propagation speed of galileon fluctuations about the spherically-symmetric background for various relative strengths of the cubic and quartic terms.  Color coding is as in Fig.~\ref{fig:metric}.}
\end{figure}

In Fig.~\ref{fig:metric} we plot the components of the effective metric for a range of different values of $x$ in the allowed range, and in Fig.~\ref{fig:speed} we demonstrate the effects on the radial and angular sound speeds for these same values. As is well-known for the galileons, radial waves propagate superluminally; note that larger quartic interaction strength correlates with faster maximum propagation speed.  It is unclear whether such superluminality presents a problem for the viability of the theory; \cite{Adams:2006sv} argues that it is an indicator that the theory cannot be UV completed and is macroscopically non-local, whereas \cite{Burrage:2011cr,Bruneton:2006gf,Babichev:2007dw,Kang:2007vs,ArmendarizPicon:2005nz} argue that causality is preserved despite superluminal signal propagation and hence the theory is no less safe than GR under the Hawking Chronology Protection Principle.  Additionally, \cite{deRham:2013hsa} shows that for a specific choice of parameters, the galileon theory (with accompanying Vainshtein mechanism and superluminal signals) is dual to a free field theory, and thus has an analytic S-matrix and is causal.

 It is also apparent from Fig.~\ref{fig:speed} that near the large central mass, the angular galileon modes propagate highly \emph{sub}luminally for galileon theories with a non-zero quartic term.  As discussed in \cite{nicolisgalileon}, this limits the validity of the static approximation we have made: the static limit is valid in the regime where the galileon propagation speed is much faster than the speed with which astrophysical objects move. Clearly then, near the Sun for example, it is necessary to solve the fully time-dependent system.

While explicitly solving~(\ref{eq:Rldiffeq}) is not possible for general parameter choices, the special values for which we can make progress are when $x$ lies at the edges of its allowed range: $x=0, 2/3$. The choice $x=0$ is the already-solved case of only cubic interactions ($r_{34}\to 0$). To explore the new phenomena exhibited when the quartic term is present, we therefore consider the choice of maximal quartic term, $x=\frac{2}{3}$.  First, however, we review the results for cubic galileons.

%~~~~~~~~~~~~~~~~~~~~~~~~~~~~~~~~~~~~~~~~~~~~~~~~~~~~~~~~~~~
\subsubsection{Cubic Galileons\label{sss:cubic}} 

Our first step is to find the (static) cubic galileon propagator - the Green's function of the wave equation linearized about $\pb$. For $\alpha_4 = \alpha_5 = 0$, the terms in \eqref{KineticTerms} become
\begin{align}
S_\text{kin} &= \int d^4x\, \left[ \phi \left( \frac{2}{\Lambda^3} \Box \bar{\pi} \Box - \frac{2}{\Lambda^3} \nabla^\mu \nabla^\nu \bar{\pi} \nabla_\mu \nabla_\nu + \half \Box \right) \phi \right]   \\
&= \int d^4x \left[ \left( 3 + \frac{2}{\Lambda^3} \Box \bar{\pi} \right) (\partial_t\phi)^2 - \left( 3 + \frac{4}{\Lambda^3} \frac{\bar{\pi}'}{r} \right) (\partial_r\phi)^2 - \left( 3 + \frac{2}{\Lambda^3}(\bar{\pi}'' + \frac{\bar{\pi}'}{r}) \right) (\partial_\Omega\phi)^2 \right] \, , \nonumber
\label{eq:new_S_kin}
\end{align}
with the background $\pb$ obtained by solving the quadratic equation in \eqref{BackgroundEquation} with the boundary condition $\pb'(r \to \infty) = 0$:
\begin{equation}
\frac{\bar{\pi}'}{r} = \frac{\alpha_2 \Lambda^3}{4 \alpha_3} \left(-1 + \sqrt{1+\frac{2}{\pi} \left(\frac{r_{23}}{r}\right)^3}\right)\, .
\label{eq:background}
\end{equation}
This equation can be integrated to obtain
\begin{equation}
\label{eq:0background}
\bar{\pi}(z) = -\left(\frac{2}{\pi}\right)^{2/3} \frac{M}{8\alpha_2 \mpl r_{23}} \left[ \frac{\Gamma\left(\frac{1}{3}\right)^2}{2^{1/3} \Gamma\left(\frac{2}{3}\right)} + \sqrt{z} \left(z^{3/2} - 4\, {}_2F_1\left(-\half,\frac{1}{6},\frac{7}{6}, -z^3\right) \right) \right]\, ,
\end{equation}
in terms of the dimensionless radial variable $z\equiv \left(\frac{\pi}{2}\right)^{1/3} \frac{r}{r_{23}}$.

As we shall now see, viewing $\phi$ as a massless scalar propagating in a curved geometry makes some of the relevant questions conceptually clear. Firstly, it is known that -- see, for instance, \cite{curvedG_theory} -- signals in curved space do not propagate solely on the light cone; there is also a tail of signals propagating everywhere within the light cone. Secondly, it makes superluminal propagation of fluctuations inside the Vainshtein radius manifest. 

For the case at hand, using the solution~(\ref{eq:background}), the relevant functions are easily read off, and in the limits $r\gg r_{23}$ (outside Vainshtein) and $r\ll r_{23}$ (inside Vainshtein), they are given by the approximate expressions 
\begin{eqnarray}
K_t(r) & \sim & \left\{\begin{array}{lr} \alpha_2  & \ \ \ \ r\gg r_{23} \\
\frac{3}{4} \alpha_2 z^{-3/2} & \ \ \ \ r\ll r_{23} \end{array}\right. \ , \nonumber \\
K_r(r) & \sim & \left\{\begin{array}{lr} \alpha_2  & \ \ \ \ r\gg r_{23} \\
\alpha_2 z^{-3/2} & \ \ \ \ r\ll r_{23} \end{array}\right. \ , \\
K_\Omega(r) & \sim & \left\{\begin{array}{lr} \alpha_2  & \ \ \ \ r\gg r_{23} \\
\frac{1}{4} \alpha_2 z^{-3/2} & \ \ \ \ r\ll r_{23} \end{array}\right. \ . \nonumber
\label{tab:kinetic_functions}
\end{eqnarray}
Hence, the effective metric approaches the flat one far outside the Vainshtein radius, whereas deep inside the Vainshtein radius
\begin{equation}
\tilde{g}_{\mu\nu} = \sqrt{\frac{3}{2\pi}}\alpha_2 \left(\frac{r_{23}}{r}\right)^{3/2} 
\text{diag} \left( -\frac{1}{3},\, \frac{1}{4},\, r^2,\, r^2\sin^2\theta \right) \, .
\end{equation}
It is now immediately apparent that well outside the Vainshtein radius, fluctuations propagate precisely luminally, as in flat space.  Well inside the Vainshtein radius, the behavior of null geodesics tells us that
\begin{equation}
c_r^2 = -\frac{\tilde{g}_{tt}}{\tilde{g}_{rr}} = \frac{K_r}{K_t} \hspace{1cm}
c_\theta^2 = -r^2 \frac{\tilde{g}_{tt}}{\tilde{g}_{\theta\theta}} = \frac{K_\Omega}{K_t}
\label{eq:propagationspeed}
\end{equation}
and in particular, the radial fluctuations propagate superluminally ($c_r^2 = \frac{4}{3}$) and the angular fluctuations propagate subluminally ($c_\theta^2 = \frac{1}{3}$).

Although at zeroth order the propagator well outside the Vainshtein radius is simply the flat-space propagator, it is less straightforward to calculate the propagator well inside the Vainshtein radius, for which we must solve for the Green's function of the differential operator $\tilde{g}^{\mu\nu} \tilde{\nabla}_\mu \tilde{\nabla}_\nu$, obeying (see \eqref{eq:Green's_fn})
\begin{equation}
\left(-3\partial_t^2 + 4\partial_r^2 + \frac{2}{r}\partial_r + \frac{\vec{L}^2}{r^2} \right) G\left( \vec{r}, \vec{r}' \right)  \ 
= -\frac{2}{\alpha_2} \sqrt{2\pi} \left(\frac{r}{r_{23}}\right)^{3/2} \delta^3\left( \vec{r}-\vec{r}' \right) \, .
\label{eq:r_Green's_fn}
\end{equation}
For simplicity we will focus on the static propagator. As already alluded to, this treatment assumes that the static force law is a good description of the dynamics despite the existence of galileon wave tails propagating inside, not on, the light-cone. Strictly speaking this assumption needs to be checked via an actual calculation. However, we argue, on physical grounds, that this is a valid assumption as long as the speed of propagation of galileons is much larger than the velocity of the astrophysical objects whose dynamics we wish to study.

We begin by defining a new variable 
\begin{equation}
\vec{\rho} = \sqrt{z} \hat{r} \, ,
\end{equation}
where $\hat{r} \equiv \vec{r}/|\vec{r}|$. In terms of this variable, we now define a re-scaled Green's function
\begin{equation}
g(\vec{\rho},\vec{\rho}') = \frac{G(\vec{r},\vec{r}')}{\rho \rho'}  \ , \qquad \rho \equiv |\vec{\rho}|\, ,
\end{equation}
which we find obeys Poisson's equation for a point mass in the new variable $\rho$; namely $-\delta^{ij} (\partial/\partial \rho^i) (\partial/\partial \rho^j) G = \delta^{3}(\vec{r}-\vec{r}')$. Thus, the rescaled $g(\rho,\rho')$ has the standard flat-space propagator solution and therefore the galileon propagator 
\begin{equation}
\langle \phi(r) \phi(r') \rangle_\text{static} = i \delta(t-t') G(r,r')
\end{equation}
must contain the term
\begin{equation}
\left(\frac{\pi}{2}\right)^{1/3} \frac{i \delta(t-t')}{2\pi \alpha_2 r_{23}} \frac{\rho \rho'}{|\vec{\rho} - \vec{\rho}'|}\, .
\label{eq:phiphi_prop}
\end{equation}

As discussed in detail in~\cite{galileonwaves}, the full propagator also must include homogeneous solutions to impose the boundary condition that when one of the endpoints is taken to coincide with the central large mass, the propagator merely renormalizes the mass in the background solution.  Taking this into account we have~\cite{galileonwaves}
\begin{equation}
\langle \phi(r) \phi(r') \rangle = \left(\frac{\pi}{2}\right)^{1/3} \frac{i \delta(t-t')}{2\pi \alpha_2 r_{23}} 
\left[ \frac{\rho \rho'}{|\vec{\rho} - \vec{\rho}'|} - \rho - \rho' + \frac{\Gamma\left(\frac{1}{6}\right) \Gamma\left(\frac{1}{3}\right)}{6\sqrt{\pi}} \right] \, .
\label{eq:phiphi_prop_full}
\end{equation}
Although it is extremely useful to have a closed form solution for this propagator, we will see in the remainder of the paper that calculating the dynamics of objects subject to this force beyond first order is technically quite difficult.

For completeness we also record here the exact solution derived in \cite{galileonwaves}:
\begin{align}
\label{GreensFunctionResult_Cubic}
G_3\left( \vec{r},\vec{r}' \right)
&= \left(\frac{\pi}{2}\right)^{1/3} \frac{1}{4\pi \alpha_2 r_{23}} \left(  
\frac{\Gamma\left[\frac{1}{3}\right] \Gamma\left[\frac{1}{6}\right]}{3\sqrt{\pi}}
- 2\sqrt{z_>} \,_2F_1\left[\frac{1}{6},\frac{1}{2};\frac{7}{6};-z_>^3\right] 
\right) \nonumber\\
&+ \left(\frac{\pi}{2}\right)^{1/3} \frac{1}{\alpha_2 r_{23}} \sum_{\ell = 1}^\infty \sum_{m = -\ell}^\ell \frac{ Y_\ell^m[\theta,\phi] \overline{Y_\ell^m}[\theta',\phi'] }{ 2\ell+1 }
z_<^{\frac{\ell+1}{2}} \ _2F_1\left[ \frac{1}{6}-\frac{\ell}{6}, \frac{1}{2}+\frac{\ell}{2}; \frac{7}{6}+\frac{\ell}{3}; -z_<^3 \right] \nonumber\\
&\times \Bigg( 
2 z_>^{-\frac{\ell}{2}} \ _2F_1\left[ \frac{\ell}{6}+\frac{1}{3}, -\frac{\ell}{2}; \frac{5}{6}-\frac{\ell}{3}; -z_>^3 \right] \\
&\qquad \qquad \qquad \qquad
+ \frac{\ell!\, \Gamma\left[ -\frac{1}{6}(2\ell+1) \right]}{\sqrt{\pi}\, \Gamma\left[\frac{1}{3}(2\ell+1)\right]} z_>^{\frac{\ell+1}{2}} \ _2F_1\left[ \frac{1}{6}-\frac{\ell}{6}, \frac{1}{2}+\frac{\ell}{2}; \frac{7}{6}+\frac{\ell}{3}; -z_>^3 \right] 
\Bigg) \, . \nonumber
\end{align}

%~~~~~~~~~~~~~~~~~~~~~~~~~~~~~~~~~~~~~~~~~~~~~~~~~~~~~~~~~~~
\subsubsection{Maximally Quartic Galileons\label{sss:maximalquartic}}

Choosing now the $x=\frac{2}{3}$ case, the background solution then obeys the equation 
\begin{equation}
\label{BackgroundPi_MaximallyQuartic}
\frac{\bar{\pi}'}{r} = \frac{\alpha_2 \Lambda^3}{2 \alpha_3} \left( -1 + \sqrt[3]{1+\frac{3}{2\pi}\left(\frac{r_{23}}{r}\right)^3} \right)  \, .
\end{equation}
 This expression can be integrated to yield
\begin{equation}
\label{eq:2/3background}
\bar{\pi}(\tilde{z}) = -\left(\frac{3}{2\pi}\right)^{2/3} \frac{M}{4\alpha_2 \mpl r_{23}} \left[ \frac{\Gamma \left(\frac{1}{3}\right)^2}{3 \Gamma \left(\frac{2}{3}\right)} 
+ \tilde{z} \left(\tilde{z}-2 \, _2F_1\left(-\frac{1}{3},\frac{1}{3};\frac{4}{3};-\tilde{z}^3\right)\right)\right] \, ,
\end{equation}
where we have chosen the integration constant to impose $\bar{\pi}(r\to\infty) = 0$ and introduced the new variable $\tilde{z} \equiv \left(\dfrac{2\pi}{3}\right)^{1/3} \dfrac{r}{r_{23}}$.  Note that this $\tilde{z}$ is a rescaling of $z$ defined for the $x=0$ case.

The functions $K_i$ which appear in the static Green's function equation are, in terms of our new variable $\tilde{z}$,
\begin{equation}
K_r(\tilde{z}) = \frac{\alpha_2}{\tilde{z}^2} (1+\tilde{z}^3)^{2/3}\,,
\hspace{3cm}
K_\Omega(\tilde{z}) = \frac{\alpha_2 \tilde{z}}{(1+\tilde{z}^3)^{1/3}} \, .
\end{equation}
Substituting these into~(\ref{eq:Rldiffeq}) it is then possible to obtain two independent solutions
\begin{eqnarray}
R_1(\tilde{z}) &=& {}_2F_1\left(\frac{\ell+1}{3}, -\frac{\ell}{3}, \frac{2}{3}, -\tilde{z}^3\right) \nonumber \\
R_2(\tilde{z}) &=& \tilde{z}\, _2F_1\left(\frac{1-\ell}{3}, \frac{\ell+2}{3}, \frac{4}{3}, -\tilde{z}^3\right) \, ,
\end{eqnarray}
so that the full radial dependence of the Green's function for each mode $\ell$ takes the form
\begin{equation}
\label{g_ell}
g_\ell(\tilde{z},\tilde{z}') \equiv R_{>,\ell}(\tilde{z}_>) R_{<,\ell}(\tilde{z}_<)
= \bigg( R_1(\tilde{z}_>) + C_{2>} R_2(\tilde{z}_>) \bigg) \bigg( C_{1<} R_1(\tilde{z}_<) + C_{2<} R_2(\tilde{z}_<) \bigg) \, .
\end{equation}

In this form we have already imposed continuity at $r=r'$. To fix the constants $C_{2>}$, $C_{1<}$ and $C_{2<}$, we shall need to impose the following boundary conditions. 

{\it Discontinuous first derivative} \qquad Firstly, the first derivative of $g_\ell(\tilde{z},\tilde{z}')$ with respect to $\tilde{z}$ needs to be discontinuous at $\tilde{z}=\tilde{z}'$, so that its second derivative gives us the needed $\delta(\tilde{z}-\tilde{z}')$. Specifically, by integrating $\widetilde{\Box} G(\tilde{z},\tilde{z}') = \delta^{(3)}(\tilde{z}-\tilde{z}')/\sqrt[4]{|g(\tilde{z}) g(\tilde{z}')|}$ with respect to $\tilde{z}$ about the point $\tilde{z}=\tilde{z}'$, one may show that
\begin{equation}
\label{WronskianI}
r'^2 K_r(r') \left(\partial_{r_>} - \partial_{r_<}\right) \big(R_\ell(r) R_\ell(r') \big)\big|_{r=r'} = -1  \, .
\end{equation}
At this point,
\begin{equation}
\label{WronskianII}
C_{2<} = C_{1<} C_{2>} + \left.\left(\frac{2\pi}{3}\right)^{1/3} \frac{1}{\alpha_2 r_{23} (1+\tilde{z}'^3)^{2/3} (R_1 \partial_{\tilde{z}} R_2 - \partial_{\tilde{z}} R_1 R_2)
} \right\vert_{\tilde{z}=\tilde{z}'}\, .
\end{equation}
Note that the second term of Eq. \eqref{WronskianII} is a constant by the equations of motion obeyed by the radial mode functions $R_i$ -- the differential equation obeyed by $q_i(\tilde{z}) \equiv (1+\tilde{z}^3)^{1/3} R_i(\tilde{z})$ contains no first derivative terms. In fact, we have normalized $R_i$ such that $(1+\tilde{z}'^3)^{2/3} (R_1 \partial_{\tilde{z}} R_2 - \partial_{\tilde{z}} R_1 R_2) = 1$. Hence
\begin{equation}
C_{2<} = C_{1<} C_{2>} + \left(\frac{2\pi}{3}\right)^{1/3} \frac{1}{\alpha_2 r_{23}}\, .
\end{equation}

{\it Observer at infinity} \qquad Next, for fixed $\tilde{z}_<$, the static Green's function must vanish as $\tilde{z}_> \to \infty$, for all $\ell \geq 0$. Using the identity 
\begin{align}
\label{2F1_zTo1/z}
\,_2F_1(\alpha,\beta;\gamma;\tilde{z})
&= \frac{\Gamma[\gamma]\Gamma[\beta-\alpha]}{\Gamma[\beta]\Gamma[\gamma-\alpha]} (-\tilde{z})^{-\alpha} \ _2F_1\left(\alpha,\alpha+1-\gamma;\alpha+1-\beta;\frac{1}{\tilde{z}}\right) \\
&\qquad \qquad
+ \frac{\Gamma[\gamma]\Gamma[\alpha-\beta]}{\Gamma[\alpha]\Gamma[\gamma-\beta]} (-\tilde{z})^{-\beta} \ _2F_1\left(\beta,\beta+1-\gamma;\beta+1-\alpha;\frac{1}{\tilde{z}}\right) \nonumber
\end{align}
on $g_\ell(\tilde{z},\tilde{z}')$ in Eq. \eqref{g_ell} results in (for $\tilde{z}\gg 1$) two terms of the form $\widetilde{C}_1 \tilde{z}_>^\ell$ and $\widetilde{C}_2 \tilde{z}_>^{-\ell-1}$. These are the familiar flat-space radial solutions.
Requiring that $\lim_{\tilde{z}_> \to \infty} g_\ell(\tilde{z},\tilde{z}') \to 0$ means $\widetilde{C}_1 = 0$, which in turn leads us to
\begin{equation}
C_{2>} = - \frac{ \Gamma\left(\frac{2}{3}\right) \Gamma\left(1+\frac{\ell}{3}\right) }{ \Gamma\left(\frac{4}{3}\right) \Gamma\left(\frac{\ell+1}{3}\right) }\, .
\end{equation}
{\it Monopole solution} \qquad The $\tilde{z}\to 0$ boundary condition we must satisfy is
\begin{equation}
\delta \bar{\pi}(\tilde{z}') = -\frac{\delta M}{\mpl} G(0,\tilde{z}')
\end{equation}
where $\delta \bar{\pi}(r')$ indicates the $\mathcal{O}(\delta M)$ part of $\bar{\pi}(M+\delta M)$. That is, since the static Green's function is the field observed at $\tilde{z}'$ produced by a static point source at $\tilde{z}$, when $\tilde{z} = 0$ this should simply amount to shifting the total mass of the point mass already present at the origin. 

Furthermore, since the background is spherically symmetric, $g_\ell(0,r')$ should only be nonzero for $\ell=0$.  This gives the condition
\begin{align}
&C_{1<}(\ell = 0) \left[1 + C_{2>}(\ell = 0) \tilde{z}' {}_2F_1\left(\frac{1}{3},\frac{2}{3},\frac{4}{3},-\tilde{z}'^3\right) \right]  \nonumber \\
&= \frac{1}{3} \left(\frac{2\pi}{3}\right)^{1/3} \frac{\Gamma\left(\frac{1}{3}\right)^2}{\Gamma\left(\frac{2}{3}\right)} \frac{1}{\alpha_2 r_{23}} \left[1 - 3 \frac{\Gamma\left(\frac{2}{3}\right)}{\Gamma\left(\frac{1}{3}\right)^2} \tilde{z}' \left( 2{}_2F_1\left(-\frac{1}{3},\frac{1}{3},\frac{4}{3},-\tilde{z}'^3\right) - \sqrt[3]{1+\tilde{z}'^3} \right) \right]\,.
\end{align}
One can check that the $\tilde{z}'$-dependent parts are equal, and thus
\begin{equation}
C_{1<}(\ell = 0) = \frac{1}{3} \left(\frac{2\pi}{3}\right)^{1/3} \frac{\Gamma\left(\frac{1}{3}\right)^2}{\Gamma\left(\frac{2}{3}\right)} \frac{1}{\alpha_2 r_{23}}\,,
\hspace{2.5cm}
C_{1<}(\ell > 0) = 0\,.
\end{equation}
{\it Result} \qquad Hence the full static propagator for the maximally quartic galileon is
\begin{align}
\label{GreensFunctionResult_Quartic}
G_4 (r,r') = 
&\left(\frac{2\pi}{3}\right)^{1/3} \frac{1}{\alpha_2 r_{23}} \Bigg\{  \frac{1}{4\pi}
\left[\frac{\Gamma \left(\frac{1}{3}\right)^2}{3 \Gamma \left(\frac{2}{3}\right)} - \tilde{z}_> {}_2F_1 \left(\frac{1}{3}, \frac{2}{3}, \frac{4}{3}, -\tilde{z}_>^3\right) \right]
\nonumber \\
&+\sum_{\ell=1}^\infty \sum_{m=-\ell}^\ell \left[ {}_2F_1 \left( -\frac{\ell}{3}, \frac{\ell+1}{3}, \frac{2}{3}, -\tilde{z}_>^3 \right) - \ell \frac{ \Gamma\left(\frac{2}{3}\right) \Gamma\left(\frac{\ell}{3}\right) }{ \Gamma\left(\frac{1}{3}\right) \Gamma\left(\frac{\ell+1}{3}\right) } \tilde{z}_> {}_2F_1 \left( \frac{1-\ell}{3}, \frac{\ell+2}{3}, \frac{4}{3}, -\tilde{z}_>^3 \right) \right]
\nonumber \\
&\qquad \qquad \qquad \qquad
	\times \tilde{z}_< {}_2F_1 \left( \frac{1-\ell}{3}, \frac{\ell+2}{3}, \frac{4}{3}, -\tilde{z}_<^3\right) Y_\ell^m (\Omega) Y_\ell^m (\Omega')^*
\Bigg\}\,.
\end{align}
Unlike in the cubic case, we did not find a closed-form Green's function in the near-source limit.

%~~~~~~~~~~~~~~~~~~~~~~~~~~~~~~~~~~~~~~~~~~~~~~~~~~~~~~~~~~~
\subsubsection{Generic Quartic Term from Perturbation Theory\label{sss:genericquartic}}

By exploiting the effective metric picture, we now adopt reasoning similar to that leading up to equations (53) and (56) of \cite{curvedG} to illustrate how one may perturbatively solve the static Green's function near the points $x=0$ and $x=2/3$, using the exact solutions we have obtained there. Let us set $x = \epsilon$ and $x = 2/3 - \epsilon$ (near $x=0$ and $x=2/3$ respectively), where $0 < \epsilon \ll 1$. Then the effective metric in \eqref{eq:g_eff_general} may be written as a power series in $\epsilon$, namely
\begin{align}
\tilde{g}_{\mu\nu} = \tilde{\bar{g}}_{\mu\nu} + \epsilon h_{\mu\nu} + \mathcal{O}(\epsilon^2)\, ,
\end{align}
where $\tilde{\bar{g}}_{\mu\nu}$ is the effective metric for either $x=0$ or $x=2/3$. Denote the exact Green's function solutions at either $x=0$ or $x=2/3$ as $\bar{G}(\vec{r}_1, \vec{r}_2)$ and the solution for arbitrary $x = \epsilon$ or $x=2/3-\epsilon$ as $G(\vec{r}_1,\vec{r}_2)$. First we consider the following integral
\begin{align}
\label{TotalDerivativeIntegral}
\int d^3 r \partial_{\mu}
\left( 
|\tilde{g}|^{\frac{1}{2}} \bar{G}\left( \vec{r}_1,\vec{r} \right) \tilde{g}^{\mu\nu} \partial_{\nu} G\left( \vec{r},\vec{r}_2 \right)
- 
|\tilde{\bar{g}}|^{\frac{1}{2}} G\left( \vec{r}_1,\vec{r} \right) \tilde{\bar{g}}^{\mu\nu} \partial_{\nu} \bar{G}\left( \vec{r},\vec{r}_2 \right)
\right) \, .
\end{align}
Here, the derivatives are with respect to the 4-coordinate $r$, and all the metric-related quantities are evaluated at $\vec{r}$. Since both geometries $\tilde{g}$ and $\tilde{\bar{g}}$ are independent of time, the time derivative term $(\mu=0)$ is automatically zero. Therefore \eqref{TotalDerivativeIntegral} is really the integral of a total spatial divergence, and if we assume the Green's functions fall off sufficiently quickly at infinity \eqref{TotalDerivativeIntegral} is identically zero. On the other hand, if we carry out the $\partial_{\mu}$ derivatives, for example $\partial_{\mu}\left( |\tilde{g}|^{\frac{1}{2}} \bar{G} \tilde{\nabla}^{\mu} G \right) = |\tilde{g}|^{\frac{1}{2}} \left( \tilde{\nabla}_{\mu} \bar{G} \tilde{\nabla}^{\mu} G + \bar{G} \widetilde{\Box}_{r} G \right)$, and proceed to employ the equations $\widetilde{\overline{\Box}} \bar{G} = -\delta^3(\vec{r}_1-\vec{r}_2)/|\tilde{\bar{g}}|^{1/2}$, $\widetilde{\Box} G = -\delta^3(\vec{r}_1-\vec{r}_2)/|\tilde{g}|^{1/2}$, and the time-independence of the Green's functions, we deduce that $G$ and $\bar{G}$ obey the integral equation
\begin{align}
\label{IntegralEquationForG}
G\left( \vec{r}_1,\vec{r}_2 \right) &= \bar{G}\left( \vec{r}_1, \vec{r}_2 \right) \\
&- \int d^3 r \left(
|\tilde{g}|^{\frac{1}{2}} \tilde{g}^{ij} \partial_{i} \bar{G}\left( \vec{r}_1, \vec{r} \right) \partial_{j} G\left( \vec{r},\vec{r}_2 \right)
- 
|\tilde{\bar{g}}|^{\frac{1}{2}} \tilde{\bar{g}}^{ij} \partial_{i} G\left( \vec{r}_1, \vec{r} \right) \partial_{j} \bar{G}\left( \vec{r},\vec{r}_2 \right)
\right) \, . \nonumber
\end{align}
By iterating this integral equation and expanding the full metric in terms of the ``background metric'' $\tilde{\bar{g}}_{\alpha\beta}$ and the perturbation $h_{\alpha\beta}$, we see that up to $\mathcal{O}(\epsilon)$, the static Green's function reads
\begin{align}
G\left( \vec{r}_1, \vec{r}_2 \right) &= \bar{G}\left( \vec{r}_1, \vec{r}_2 \right) 
- \epsilon \int d^3 r |\tilde{\bar{g}}|^{\frac{1}{2}} \partial_{i} \bar{G}\left( \vec{r}_1, \vec{r} \right) 
\left( \frac{1}{2} \tilde{\bar{g}}^{\mu\nu} h_{\mu\nu} \tilde{\bar{g}}^{ij} - h^{ij} \right)
\partial_{j} \bar{G}\left( \vec{r}, \vec{r}_2 \right) \, , 
\end{align}
where the perturbation with upper indices is defined as $h^{ij} \equiv \tilde{\bar{g}}^{i\mu} \tilde{\bar{g}}^{j\nu} h_{\mu\nu}$.  We remark in passing that, by iterating \eqref{IntegralEquationForG} repeatedly, perturbation theory can be carried out, in principle, to arbitrary order in $\epsilon$. The background Green's function is an expansion in spherical harmonics with coefficients $R_{>,\ell}(r_>) R_{<,\ell}(r_<)$, where
\begin{align}
\label{eq:xis0Rs}
&{\bf \underline {x=0:}}    \\
&R_{<,\ell=0}(r_<) = \left(\frac{\pi}{2}\right)^{1/3} \frac{1}{\alpha_2 r_{23}} \frac{\Gamma\left(\frac{1}{3}\right) \Gamma\left(\frac{7}{6}\right)}{\Gamma\left(\half\right)}          \nonumber \\
&R_{<,\ell\neq 0}(r_<) = \left(\frac{\pi}{2}\right)^{1/3} \frac{1}{\alpha_2 r_{23} (2\ell+1)}  z_<^\frac{\ell+1}{2} {}_2F_1\left(\frac{1-\ell}{6},\frac{\ell+1}{2},\frac{2\ell+7}{6},-z_<^3 \right)      \nonumber \\
&R_{>,\ell}(r_>) = 2 z_>^{-\frac{\ell}{2}} 
{}_2F_1\left( \frac{\ell+2}{6}, -\frac{\ell}{2}, \frac{5-2\ell}{6}, -z_>^3 \right) + 
\frac{\ell!\, \Gamma\left(-\frac{1}{6}(2\ell+1)\right)}{\sqrt{\pi}\, \Gamma\left(\frac{1}{3}(2\ell+1)\right)}
z_>^\frac{\ell+1}{2} {}_2F_1\left( \frac{1-\ell}{6}, \frac{\ell+1}{2}, \frac{2\ell+7}{6}, -z_>^3 \right)   \nonumber  \\
\label{eq:xis2/3Rs}
&{\bf \underline {x=\frac{2}{3}:}}    \\
&R_{<,\ell=0}(r_<) = \left(\frac{2\pi}{3}\right)^{1/3} \frac{1}{\alpha_2 r_{23}} \frac{\Gamma\left(\frac{1}{3}\right) \Gamma\left(\frac{4}{3}\right)}{\Gamma\left(\frac{2}{3}\right)}     \nonumber \\
&R_{<,\ell\neq 0}(r_<) = \left(\frac{2\pi}{3}\right)^{1/3} \frac{1}{\alpha_2 r_{23}} \tilde{z}_<\, {}_2F_1\left(\frac{1-\ell}{3}, \frac{\ell+2}{3}, \frac{4}{3}, -\tilde{z}_<^3 \right)                  \nonumber \\
&R_{>,\ell}(r_>) = {}_2F_1\left(-\frac{\ell}{3}, \frac{\ell+1}{3}, \frac{2}{3}, -\tilde{z}_>^3\right) - \frac{\Gamma\left(\frac{2}{3}\right) \Gamma\left(1+\frac{\ell}{3}\right)}{\Gamma\left(\frac{4}{3}\right) \Gamma\left(\frac{\ell+1}{3}\right)} \tilde{z}_> {}_2F_1\left( \frac{1-\ell}{3}, \frac{\ell+2}{3}, \frac{4}{3}, -\tilde{z}_>^3\right) \, .     \nonumber
\end{align}

Via the orthonormality of the spherical harmonics, integrating-by-parts the angular derivatives, using the eigenvalue equation for the $Y_\ell^m$s, and changing the radial integration variable to $z$ (or $\tilde{z}$), we arrive at the formula
\begin{align}
\label{GreensFunction_FirstOrderFormula}
&G\left( \vec{r}_1, \vec{r}_2 \right) = \bar{G}\left( \vec{r}_1, \vec{r}_2 \right) 
- \epsilon \sum_{\ell,m} Y_\ell^m(\Omega_1) Y_\ell^m(\Omega_2)^* \\
&\times \left( R_{>,\ell}(r_1) R_{>,\ell}(r_2) I_{>>,\ell} + R_{>,\ell}(r_>) R_{<,\ell}(r_<) I_{><,\ell} + R_{<,\ell}(r_1) R_{<,\ell}(r_2) I_{<<,\ell} \right) 
+\mathcal{O}(\epsilon^2) \, . \nonumber
\end{align}
Only one-dimensional integrals remain. Defining $M^{(r)}$ to be the radial $ij = rr$ and $\Omega^{AB} M^{(\Omega)}$ to be the angular $ij = AB$ components of the (diagonal) matrix $|\tilde{\bar{g}}|^{\frac{1}{2}} \left( \frac{1}{2} \tilde{\bar{g}}^{\mu\nu} h_{\mu\nu} \tilde{\bar{g}}^{ij} - h^{ij} \right)$, where $\Omega^{AB} = \text{diag}(1,1/\sin^2\theta)$ is the inverse metric on a 2-sphere, we find that
\begin{align}
\label{I_>>}
I_{>>,\ell} \equiv \int_0^{z_<} d z \left[ \left( \partial_{z} R_{<,\ell}(z) \right)^2 M^{(r)}(z) + \ell(\ell+1) \left( R_{<,\ell}(z) \right)^2 M^{(\Omega)}(z) \right]
\end{align}
\begin{align}
\label{I_<>}
I_{><,\ell} \equiv \int_{z_<}^{z_>} d z \left[ \partial_{z} R_{>,\ell}(z) M^{(r)}(z) \partial_{z} R_{<,\ell}(z) + \ell(\ell+1) R_{>,\ell}(z) M^{(\Omega)}(z) R_{<,\ell}(z) \right]
\end{align}
\begin{align}
\label{I_<<}
I_{<<,\ell} \equiv \int_{z_>}^\infty d z \left[ \left( \partial_{z} R_{>,\ell}(z) \right)^2 M^{(r)}(z) + \ell(\ell+1) \left( R_{>,\ell}(z) \right)^2 M^{(\Omega)}(z) \right] \, .
\end{align}
For $x=\epsilon$, we have $z \equiv \left(\frac{\pi}{2}\right)^{1/3} \frac{r}{r_{23}}$ and
\begin{align}
M^{(r)}(z) 			&= \frac{\alpha_2 r_{23}}{6(4\pi)^{1/3}} \sqrt{\frac{z}{1+z^3}} \left[ 2\left(1+z^3\right) \left(\sqrt{1+z^{-3}} - 1\right) -1 \right]  \\
M^{(\Omega)}(z) &= \frac{\alpha_2 r_{23}}{6 (4\pi)^{1/3}} \frac{1}{\sqrt{1+z^{-3}}} \left[\left(2-z^{-3}\right)\sqrt{1+z^{-3}} - \frac{3}{4z^3 (1+z^3)} - 2 \right] \, .
\end{align}
For $x = 2/3-\epsilon$, we have $\tilde{z} \equiv \left(\frac{2\pi}{3}\right)^{1/3} \frac{r}{r_{23}}$ and
\begin{align}
M^{(r)}\left( \tilde{z} \right) 		&= \frac{\alpha_2 r_{23}}{(18\pi)^{1/3}} \frac{1}{\sqrt[3]{1+\tilde{z}^3}} \left[3\tilde{z}^3 \left(\sqrt[3]{1+\tilde{z}^{-3}} - 1\right) -1 \right]  \\
M^{(\Omega)}\left( \tilde{z} \right) 	&= \frac{\alpha_2 r_{23}}{(18\pi)^{1/3}} \frac{1}{\sqrt[3]{1+\tilde{z}^{-3}}\left(1+\tilde{z}^3\right)} \left[ 3 \left(1+\tilde{z}^3\right) \left(\sqrt[3]{1+\tilde{z}^{-3}} -1 \right) - 1\right]  \, .
\end{align}
For $\ell=0$, the integrals \eqref{I_<<}, \eqref{I_<>} and \eqref{I_>>} can be evaluated exactly. Referring to \eqref{eq:xis0Rs} and \eqref{eq:xis2/3Rs}, we see that $R_{<,0}$ is a constant for both $x = 0,2/3$. This means the only non-zero integral is $I_{<<,\ell=0}$, which is given by
\begin{align}
\label{I<<0_IofII}
I_{<<,\ell=0}(x\approx 0) &= \left(\frac{2}{\pi}\right)^{1/3} \frac{\alpha_2 r_{23}}{18} \left( \sqrt{\frac{z_>}{1+z_>^3}} + \frac{3}{z_>} - 4\frac{\Gamma\left(\frac{1}{3}\right) \Gamma\left(\frac{4}{3}\right)}{2^{1/3} \Gamma\left(\frac{2}{3}\right)} R_{>,0}(z_>) \right)
\\
\label{I<<0_IIofII}
I_{<<,\ell=0}\left(x\approx 2/3\right) &= \left(\frac{3}{2\pi}\right)^{1/3} \frac{\alpha_2 r_{23}}{3} \left( \frac{3}{\sqrt[3]{1+\tilde{z}_>^3}} - \frac{\tilde{z}_>}{\sqrt[3]{\left(1+\tilde{z}_>^3\right)^2}} - 2\frac{\Gamma\left(\frac{1}{3}\right) \Gamma\left(\frac{4}{3}\right)}{\Gamma\left(\frac{2}{3}\right)} R_{>,0}(\tilde{z}_>) \right) \, .
\end{align}

These expressions enable us to perform a check on the Green's function perturbation theory: the $\ell=0$ piece of the static Green's function, when $r_<\to0$, should now correspond to the $\mathcal{O}(\epsilon)$-accurate coefficient of $\delta M/\mpl$ of the background solution $\bar{\pi}$ of the central mass, upon shifting it by $M \to M + \delta M$. One may confirm this by inserting in \eqref{GreensFunction_FirstOrderFormula} the expressions in \eqref{eq:xis0Rs}, \eqref{eq:xis2/3Rs}, \eqref{I<<0_IofII} and \eqref{I<<0_IIofII}; and comparing the result with the following expressions for $\bar{\pi}$. When $x=\epsilon$, the $\mathcal{O}(\epsilon)$ correction to Eq.~(\ref{eq:0background}) is given by
\begin{align}
\bar{\pi}(z) = &\bar{\pi}_{x=0}(z) - \epsilon \left(\frac{2}{\pi}\right)^{2/3} \frac{M}{36 \alpha_2 \mpl r_{23}} \Bigg[ \frac{\Gamma\left(\frac{1}{3}\right)^2}{2^{1/3} \Gamma\left(\frac{2}{3}\right)}   \nonumber \\
&+ \sqrt{z} \left( \frac{3}{2} z^{3/2} - \half \sqrt{1+z^3} - 4\,{}_2F_1\left(-\half,\frac{1}{6},\frac{7}{6},-z^3\right) \right) - \frac{3}{4z} \Bigg] \, .
\end{align}
Similarly, when $x=\frac{2}{3}-\epsilon$, the $\mathcal{O}(\epsilon)$ correction to Eq.~\ref{eq:2/3background} is given by
\begin{equation}
\bar{\pi}(\tilde{z}) = \bar{\pi}_{x=2/3}(\tilde{z}) - \epsilon \left(\frac{3}{2\pi}\right)^{2/3} \frac{M}{6\alpha_2 \mpl r_{23}} \left[ \frac{\Gamma\left(\frac{1}{3}\right)^2}{3\Gamma\left(\frac{2}{3}\right)} + \tilde{z} \left( \frac{3}{2} \tilde{z} - {}_2F_1\left(\frac{1}{3},\frac{2}{3},\frac{4}{3},-\tilde{z}^3\right) \right) - \frac{3}{2} \left(1+\tilde{z}^3\right)^{2/3} \right] \, .
\end{equation}

For $\ell \geq 1$, the integrals in \eqref{I_>>}, \eqref{I_<>}, and \eqref{I_<<} consist of terms with products of two $\,_2 F_1$s and $r^a (1+r^3)^b$, where $a$ and $b$ are integers or rational numbers. They likely cannot be performed exactly, but it is possible that, in the limits $r_1,r_2 \ll r_{23}$ and $r_1,r_2 \gg r_{23}$ -- by writing the $\,_2 F_1$s appropriately so that they may be replaced with the first few terms of their Taylor series -- an approximate expression may be obtained. We leave these technical issues to possible future work.

%~~~~~~~~~~~~~~~~~~~~~~~~~~~~~~~~~~~~~~~~~~~~~~~~~~~~~~~~~~~~~~~~~~~~~~~~~~~~~~~~~~~~~~~~~~~~~~~~~~~~~~~~~~~~
%~~~~~~~~~~~~~~~~~~~~~~~~~~~~~~~~~~~~~~~~~~~~~~~~~~~~~~~~~~~~~~~~~~~~~~~~~~~~~~~~~~~~~~~~~~~~~~~~~~~~~~~~~~~~
\section{Nonlinearities\label{s:nonlinearities}}

Having now obtained (exactly or perturbatively, depending on the choice of parameters) the propagator for galileons in the presence of a large central mass, we wish to study perturbatively the motion of astrophysical objects subject to the full nonlinear galileon force.  We will proceed using the field theoretic effective action framework derived for GR for a two-body system in \cite{EFT} and generalized to $n$-body systems in \cite{nbody}.  For simplicity, we will henceforth restrict the analysis to only the cubic interactions ($x=0$ case).

%~~~~~~~~~~~~~~~~~~~~~~~~~~~~~~~~~~~~~~~~~~~~~~~~~~~~~~~~~~~~~~~~~~~~~~~~~~~~~~~~~~~~~~~~~~~~~~~~~~~~~~~~~~~~
\subsection{Diagrammatic Construction of Perturbation Theory\label{ss:diagrams}}

\begin{figure}[ht]
\includegraphics[width=\textwidth]{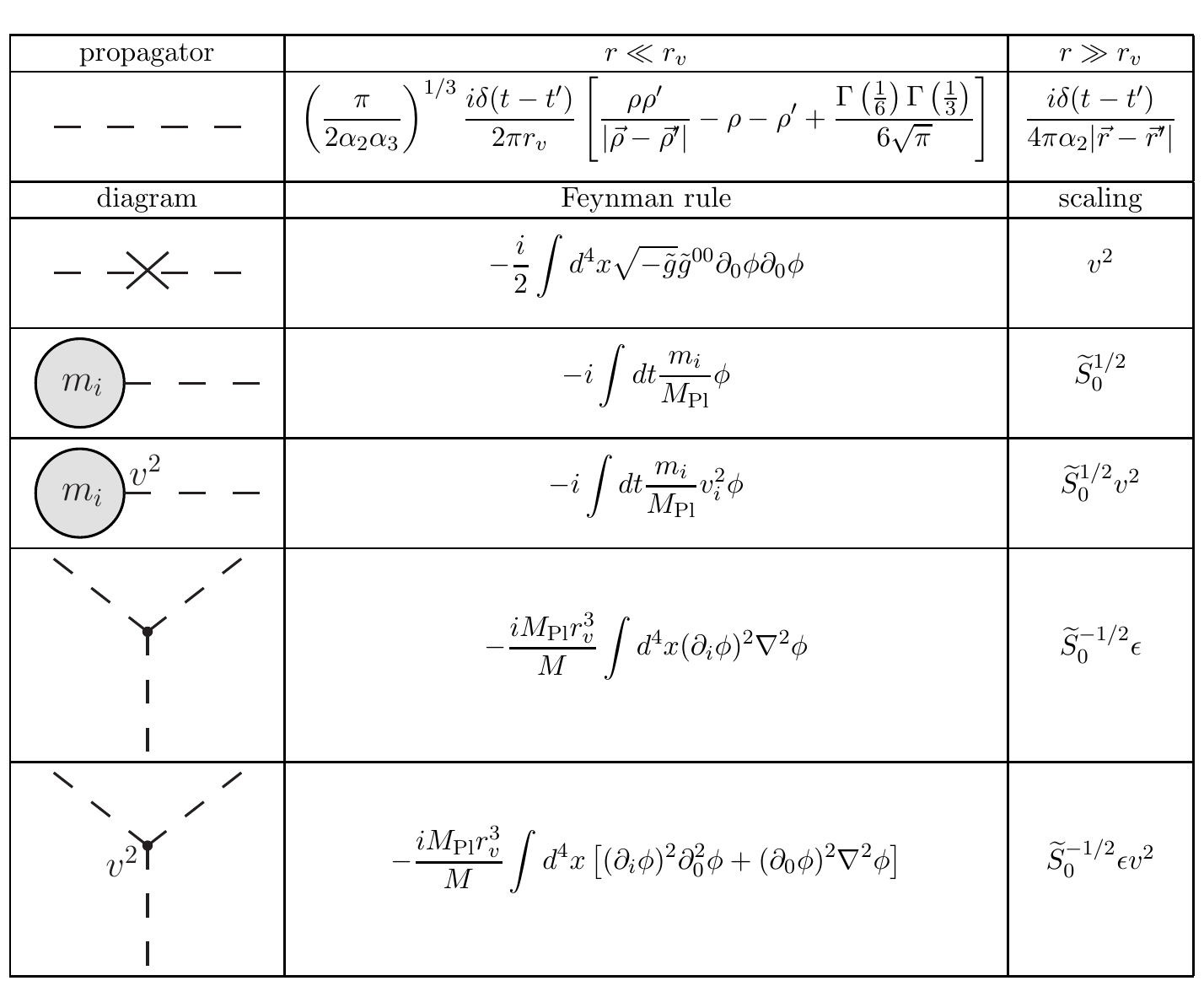}
\caption{\label{fig:diagrams} Ingredients for power counting and calculation of diagrams in the effective action.}
\end{figure}

With a closed-form static propagator well within and well outside the Vainshtein radius in hand, we now consider the construction of the perturbative expansion.  The approach we will use is familiar from quantum field theory - we will construct an effective action by integrating out $\phi = \pi - \bar{\pi}$, leaving an action dependent only on the positions and velocities of the point particles:
\begin{equation}
e^{i S_\text{eff}[\vec{x}_a,\vec{v}_a]} = \int \mathcal{D}\phi\, e^{i S[\phi,\vec{x}_a,\vec{v}_a]} \, .
\end{equation}
The functional integral can be performed up to an irrelevant overall factor $\mathcal{N}$ by
\begin{equation}
e^{i S_\text{eff}[\vec{x}_a,\vec{v}_a]}
= \left. \mathcal{N} \text{exp}\left[i S_\text{int}\left[\frac{1}{i}\frac{\delta}{\delta J}\right]\right] 
\text{exp}\left[-\half \int d^4x d^4y J(x) \langle \phi(x) \phi(y) \rangle J(y) \right] \right|_{J=0} \, ,
\end{equation}
where the interaction part of the action $S_\text{int}\left[\frac{1}{i}\frac{\delta}{\delta J}\right]$ is defined as $S_\text{int} = S - S_\text{kin}$, with all occurrences of $\phi$ replaced by $\frac{1}{i}\frac{\delta}{\delta J}$.  Expanding the interaction exponential then leads to an infinite series of terms, which can be represented graphically as Feynman diagrams.  The dictionary of what each graphical object means mathematically is given in Fig.~\ref{fig:diagrams}.  Complete diagrams are created by Wick contracting each occurrence of $\phi$ with a corresponding occurrence of $\phi$ in another diagram piece.  Note that, since we are interested in the classical galileon force, we neglect all loop diagrams and consider only tree-level diagrams.  This procedure is completely equivalent to the Born approximation, and in particular, one need not worry about properly defining the measure for the path integral for non-canonical field theories.

The classical effective action is thus given by 
\begin{equation}
e^{i S_\text{eff}} = e^{\sum (\text{fully-connected tree-level diagrams})} \, .
\label{diagramsum}
\end{equation}
It is important to understand how each Feynman diagram scales with the physical scales of the problem at hand, so that we may identify the relevant expansion parameters to organize our calculations and also understand the domain of their validity. We begin by noting that, since the galileon only appears as a propagator, it is appropriate to consider it to scale as the square root of the propagator in \eqref{eq:phiphi_prop_full}
\begin{equation}
\phi \sim \sqrt{\langle \phi\phi \rangle} \sim \frac{v^{1/2}}{r_v^{3/4} r^{1/4}} \, .
\label{eq:propscaling}
\end{equation}

Were we to be interested in the results of our calculation in a world without a spin-2 graviton, it would be necessary to understand the analogue of the virial theorem for a purely scalar force. Technically, the virial theorem should be derived by requiring that the kinetic energy of the test masses scale in the same way as the one-graviton exchange (potential) between the large central mass and a test mass.  Thus, the purely galileon virial velocity is derived as
\begin{equation}
S_0 \equiv m v_{\rm gal} r \sim \frac{m M}{\mpl^2 v_{\rm gal}} \left(\frac{r}{r_v} \right)^{3/2}   \hspace{1cm} \Rightarrow \hspace{1cm}
v_{\rm gal}^2 \sim \frac{M}{\mpl^2 r} \left(\frac{r}{r_v}\right)^{3/2} \, .
\label{eq:galileonvirial}
\end{equation}
\\
However, we are obviously motivated by the eventual goal of including gravity, and so it is appropriate here to invoke the standard GR virial theorem where necessary.
\begin{equation}
v_{\rm GR}^2 \sim \frac{M}{\mpl^2 r} \, .
\label{eq:GRvirial}
\end{equation}
In the case of test masses well outside the Vainshtein radius, the virial theorem for the galileon case matches that of the GR case, and the propagator is the standard flat-space one.

The diagrams that result, what they represent, and how they scale under both versions of the virial theorem are collected in Fig.~\ref{fig:diagrams} for both inside and outside the Vainshtein radius.
The effective action is given by the sum of diagrams that can be constructed from these basic pieces, and arranges itself into an expansion in two parameters:
\begin{equation}
S_\text{eff} = \widetilde{S}_0 (1 + v^2 + \cdots) (1 + \epsilon + \cdots) \, ,
\label{eq:Seff}
\end{equation}
with the values of $\widetilde{S}_0$ and $\epsilon$ depending on the choice of virial theorem and on the distance regime via
\begin{equation}
\hfill
\widetilde{S}_0 =
\begin{cases}
S_0 \frac{m}{M}  &r \ll r_v \text{ with } v_{\rm gal} \text{ or } r \gg r_v   \\
S_0 \frac{m}{M} \left(\frac{r}{r_v}\right)^{3/2}  &r \ll r_v \text{ with } v_{\rm GR} 
\end{cases}\, ,
\hspace{1cm}
\epsilon = 
\begin{cases}
\frac{m}{M}  &r\ll r_v  \\
\frac{m}{M} \left(\frac{r_v}{r}\right)^3  &r\gg r_v
\end{cases} \, .
\hfill
\end{equation}

Note that well within the Vainshtein radius and using the GR virial theorem, the overall amplitude of the effective action $\widetilde{S}_0$ is suppressed relative to the GR amplitude by a factor of $\left(\frac{r}{r_v}\right)^{3/2}$.  This is one manifestation of the screening mechanism within the Vainshtein radius.  Also note that the second expansion parameter is, indeed, small, given that the masses whose dynamics we are considering are small compared to that sourcing the background.  However, as we will discuss later in more detail, when distance hierarchies are taken into account this parameter can become large in some regions of interest for comparison with astrophysical observations.  This problem is not present outside the Vainshtein radius due to the additional small factor of $ \left(\frac{r_v}{r}\right)^3$.

We can now justify our consideration of galileons in flat space by arguing that, at least deep within the Vainshtein radius of the Sun, galileon-graviton interactions are suppressed relative to galileon self-interactions by multiplicative factors of $\epsilon$, $v^2$ and/or $(r/r_v)$. If we denote the graviton propagator by a double wavy line, the effective interaction between planets orbiting the Sun is, at first order in $h/\mpl$, described by the Feynman graphs:
\begin{center}
\begin{tabular}{m{0.1\textwidth} l r m{0.1\textwidth} l r m{0.1\textwidth}}
&\raisebox{-0.5\height}{\includegraphics[width=0.2\textwidth]{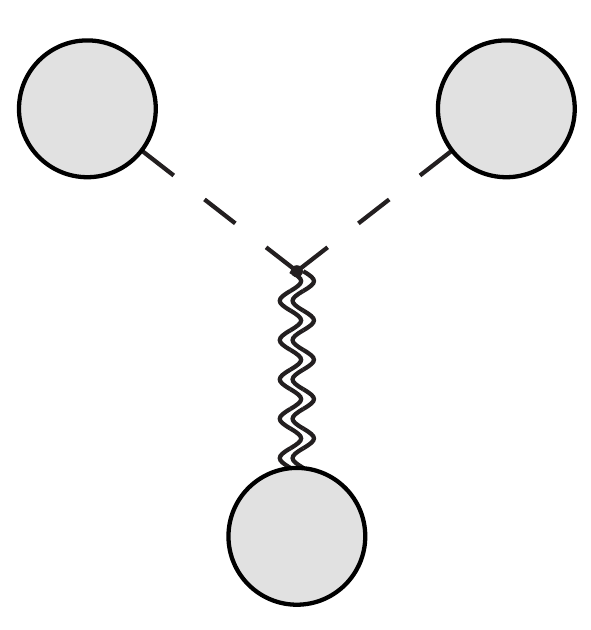}}
& $\sim \widetilde{S}_0 \epsilon v^2 \left(\dfrac{r}{r_v}\right)^{3/2}$
&
&\raisebox{-0.5\height}{\includegraphics[width=0.2\textwidth]{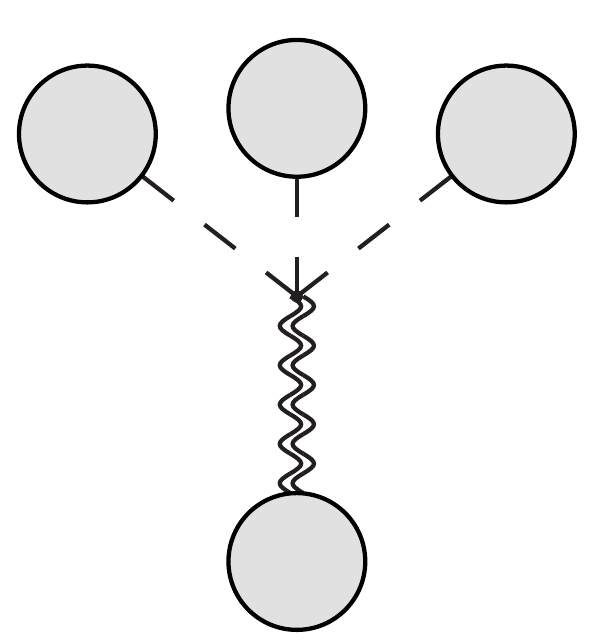}}
& $\sim \widetilde{S}_0 \epsilon^2 v^2$
&
\end{tabular}
\end{center}
Diagrams with more graviton lines would be suppressed by higher powers of $G_\text{N} m_i/r \sim v^2$. To arrive at these scaling relations, we first covariantize the action in Eq.~\eqref{eq:quinticaction}, replacing partial derivatives with covariant ones, and expand about flat spacetime: $g_{\mu\nu} = \eta_{\mu\nu} + h_{\mu\nu}/M_\text{pl}$. Then, because gravity is dominant over galileon forces here, we use the GR virial theorem. 

Additionally, we can confirm that the correction to the gravitational force of the Sun on the planets due to galileon-graviton interactions is small due to Vainshtein screening. This is done by (perturbatively) solving for $h_{\mu\nu}(x)/\mpl$ by expanding the covariantized action in Eq.~\eqref{eq:quinticaction} about flat spacetime and evaluating on the galileon background $\pi = \bar{\pi}$. 

Explicitly, the correction coming from the $(h/\mpl)(\partial \bar{\pi})^2$ interaction is given by
\begin{align}
\frac{1}{\mpl}\langle h_{\mu\nu}(x) \rangle_{\bar{\pi}\bar{\pi}} 
&= \int d^4 y \left[ \frac{\alpha_2}{2} 
\left( \eta^{\alpha\kappa} \eta^{\beta\lambda} - \half \eta^{\alpha\beta} \eta^{\kappa\lambda} \right) \langle h_{\mu\nu}(x) h_{\alpha\beta}(y) \rangle \partial_\kappa \bar{\pi}(y) \partial_\lambda \bar{\pi}(y) \right]
\\
&\sim \Psi_h \frac{\bar{\pi}}{\mpl} \left(\frac{r}{r_v}\right)^{\{3/2,2\}} 
\end{align}
for \{purely cubic, non-zero quartic\} galileon theories. Here, $\Psi_h \sim G_\text{N} M_\odot/r$ is the Newtonian gravitational potential of the Sun. Because the first order general relativistic correction to $h_{\mu\nu}(x)/\mpl$ begins at $\mathcal{O}(\Psi_h^2)$, we see that galileon-graviton interactions are suppressed as long as $\frac{\bar{\pi}}{\mpl} \left(\frac{r}{r_v}\right)^{\{3/2,2\}} \ll \Psi_h$.

Similarly, the $(h/\mpl)\partial^2 \bar{\pi} (\partial \bar{\pi})^2/\Lambda^3$ interaction term yields a correction that reads
\begin{align}
\frac{1}{\mpl}\langle h_{\mu\nu}(x) \rangle_{\bar{\pi}\bar{\pi}\bar{\pi}} 
&= \int d^4 y \Bigg[ \frac{\alpha_3}{2\Lambda^3} 
\left( \eta^{\alpha\rho} \eta^{\beta\sigma} \eta^{\kappa\lambda} + \eta^{\alpha\beta} \eta^{\kappa\rho} \eta^{\lambda\sigma} - 2 \eta^{\alpha\rho} \eta^{\kappa\beta} \eta^{\lambda\sigma} \right)  \nonumber\\
&\qquad\qquad\qquad\qquad \times\langle h_{\mu\nu}(x) h_{\alpha\beta}(y) \rangle \partial_\kappa \partial_\lambda \bar{\pi}(y) \partial_\rho \bar{\pi}(y) \partial_\sigma \bar{\pi}(y) \Bigg]
\\
& \sim \Psi_h \frac{\bar{\pi}}{\mpl} \left(\frac{r}{r_v}\right)^{\{0,1\}}
\end{align}
and the $(h/\mpl)(\partial^2 \bar{\pi})^2 (\partial \bar{\pi})^2/\Lambda^6$ interaction term (which is of course absent in the purely cubic galileon theory) gives us
\begin{align}
\frac{1}{\mpl}\langle h_{\mu\nu}(x) \rangle_{\bar{\pi}\bar{\pi}\bar{\pi}\bar{\pi}} 
= &\int d^4 y \Big[
 \left(-\frac{\alpha_4}{4\Lambda^6}\right)
\langle h_{\mu\nu}(x) h_{\alpha\beta}(y) \rangle 
\Big( M^{\alpha\beta\gamma\delta\kappa\lambda\rho\sigma} \partial_\gamma \partial_\delta \bar{\pi}(y) \partial_\rho \partial_\sigma \bar{\pi}(y) \partial_\kappa \bar{\pi}(y) \partial_\lambda \bar{\pi}(y) 
\nonumber \\
&+ N^{\alpha\beta\gamma\delta\kappa\lambda\rho\sigma} \partial_\gamma \partial_\delta \partial_\sigma \bar{\pi}(y) \partial_\kappa \bar{\pi}(y) \partial_\lambda \bar{\pi}(y) \partial_\rho \bar{\pi}(y)  \Big)
\Big]
\\
\sim &\Psi_h \frac{\bar{\pi}}{\mpl}
\, ,
\end{align}
with
\begin{align}
M^{\alpha\beta\gamma\delta\kappa\lambda\rho\sigma}
= & -2\eta^{\alpha\kappa}\eta^{\beta\lambda}\eta^{\gamma\delta}\eta^{\rho\sigma} 
- \half\eta^{\alpha\beta}\eta^{\gamma\delta}\eta^{\kappa\lambda}\eta^{\rho\sigma} 
- 2\eta^{\alpha\beta}\eta^{\gamma\delta}\eta^{\kappa\sigma}\eta^{\lambda\rho} 
+ 6\eta^{\alpha\kappa}\eta^{\beta\gamma}\eta^{\delta\lambda}\eta^{\rho\sigma}
\nonumber \\ 
&- 8\eta^{\alpha\kappa}\eta^{\beta\rho}\eta^{\gamma\lambda}\eta^{\delta\sigma} 
+ 3\eta^{\alpha\beta}\eta^{\gamma\lambda}\eta^{\delta\rho}\eta^{\kappa\sigma} 
- \half\eta^{\alpha\beta}\eta^{\gamma\sigma}\eta^{\delta\rho}\eta^{\kappa\lambda} 
+ \eta^{\alpha\gamma}\eta^{\beta\delta}\eta^{\kappa\lambda}\eta^{\rho\sigma} 
\nonumber \\
&+ 2\eta^{\alpha\gamma}\eta^{\beta\delta}\eta^{\kappa\sigma}\eta^{\rho\lambda} 
+ 3\eta^{\alpha\lambda}\eta^{\beta\kappa}\eta^{\gamma\sigma}\eta^{\delta\rho}
-2\eta^{\alpha\delta}\eta^{\beta\rho}\eta^{\gamma\lambda}\eta^{\kappa\sigma}
\\
N^{\alpha\beta\gamma\delta\kappa\lambda\rho\sigma}
= & -\eta^{\alpha\beta}\eta^{\gamma\delta}\eta^{\kappa\lambda}\eta^{\rho\sigma} 
+ \eta^{\alpha\beta}\eta^{\gamma\lambda}\eta^{\delta\kappa}\eta^{\rho\sigma} 
+ \eta^{\alpha\kappa}\eta^{\beta\lambda}\eta^{\gamma\delta}\eta^{\rho\sigma}
- \eta^{\alpha\delta}\eta^{\beta\kappa}\eta^{\gamma\lambda}\eta^{\rho\sigma}
\nonumber \\ 
&+ \eta^{\alpha\gamma}\eta^{\beta\delta}\eta^{\kappa\lambda}\eta^{\rho\sigma} \, .
\end{align}
Thus, we may consistently neglect gravity when considering galileon forces due to the Sun's galileon field $\bar{\pi}$ -- for instance, calculating perihelion precession -- and when computing the effective potential between the planetary bodies themselves, as long as these phenomenon are taking place deep within the Vainshtein radius of the Sun.

In the next section we will describe how to calculate the dynamics due to cubic galileon forces in the region where the point masses in question lie well outside the Sun's and each other's Vainshtein radii.

%~~~~~~~~~~~~~~~~~~~~~~~~~~~~~~~~~~~~~~~~~~~~~~~~~~~~~~~~~~~~~~~~~~~~~~~~~~~~~~~~~~~~~~~~~~~~~~~~~~~~~~~~~~~~
\subsection{Outside the Vainshtein Radius\label{ss:outsiderv}}

The procedure for calculating the galileon forces perturbatively is straightforward in the case where the two interacting particles are well outside the Vainshtein radius.  This is because, as expected, the galileon theory behaves as a simple scalar-tensor theory in this regime, and nonlinear interaction terms can be treated perturbatively.  Thus, using the convention
\begin{equation}
\widetilde{\Box} \langle \phi \phi \rangle = -i \frac{\delta^4 (r-r')}{\sqrt{-\tilde{g}}} \, ,
\end{equation}
the galileon propagator takes the flat-space form
\begin{equation}
\langle \phi(r)\phi(r') \rangle = \frac{i \delta(t-t')}{4\pi \alpha_2 |\vec{r}-\vec{r}'|} \, .
\end{equation}
We wish to understand how each term of our effective action scales with the physical scales of the solar system. In general relativity the effective action of the solar system arranges into an expansion in $v^2$ via $S_\text{eff} = S_0 (1 + v^2 + \cdots)$. This expansion in only powers of $v^2$ neglects the large disparity of distances and masses present in the solar system. This subtlety will be important in the galileon case, where we will find that once the planetary bodies in question get too close to each other, for example in the Sun-Earth-Moon configuration, one loses perturbative control over the effective action calculation at hand. The same issue does not present a problem in GR within the solar system, because $G_N m/|\vec{r}_i - \vec{r}_j| \ll 1$ regardless of the planet's mass $m$ and separation distance $|\vec{r}_i - \vec{r}_j|$.

Leaving aside, for the moment, the issue of the disparity of distances and masses, the power counting of diagrams in Fig.~\ref{fig:diagrams} is straightforward.  As described in \cite{EFT}, the relevant time scale in the dynamics of objects orbiting a central mass is set by their velocity, and so it is sensible to consider all factors of time to scale as $\frac{r}{v}$.  An additional subtlety to consider is the virial theorem used to organize potential terms within the expansion, as we have discussed.  

These arguments lead to the scaling given in Fig.~\ref{fig:diagrams} and to the resulting conclusion that the additional expansion parameter is $\frac{m}{M} \left(\frac{r_v}{r}\right)^3$.  Using this, we can then calculate the effective action via~(\ref{diagramsum}) as discussed above.

At the first few orders, the calculation is relatively simple. At $\mathcal{O}(\widetilde{S}_0)$, only the one-galileon-exchange diagram contributes, giving
\begin{equation}
S_\text{eff}^{(0)} = -\frac{m_1 m_2}{4\pi \alpha_2 \mpl^2} \int \frac{dt}{|\vec{r}_1-\vec{r}_2|} \, ,
\end{equation}
while at $\mathcal{O}(\widetilde{S}_0 \epsilon)$, only the three-galileon-vertex diagram is relevant, yielding
\begin{equation}
S_\text{eff}^{(\epsilon)} = \frac{\alpha_3 m_1 m_2 m_3}{8\pi^2 \alpha_2^3 \mpl^2 M} r_v^3
\int dt \left( \frac{ \hat{R}_{13}\cdot \hat{R}_{23} }{ R_{13}^2 R_{23}^2 } - \frac{ \hat{R}_{12}\cdot \hat{R}_{23} }{ R_{12}^2 R_{23}^2 } + \frac{ \hat{R}_{12}\cdot \hat{R}_{13} }{ R_{12}^2 R_{13}^2 } \right) \, ,
\end{equation}
with $\vec{R}_{ab} = \vec{r}_a - \vec{r}_b$.

We now see concretely, e.g.\ when object 2$=$object 3, that the small correction $\mathcal{L}^{(\epsilon)} \sim \frac{m_2}{M} \left(\frac{r_v}{R_{12}}\right)^3 \mathcal{L}^{(0)}$, or, in terms of the Vainshtein radius of object 2, $\mathcal{L}^{(\epsilon)} \sim \left(\frac{r_{v,2}}{R_{12}}\right)^3 \mathcal{L}^{(0)}$.  The expansion does not depend at all on the existence of the large central mass, but now it becomes clear that the relevant distance scale is the Vainshtein radius of the small masses whose dynamics we are considering.  Thus, to remain within the regime for which perturbation theory about the flat galileon background is valid, the small masses must be well-separated from each other as well as from the large central mass.

At $\mathcal{O}(\widetilde{S}_0 v^2)$ things are, in principle, a little more complicated. There are three diagrams that contribute: the $v^2$ corrections to the point-particle Lagrangian for each mass involved in the one-galileon-exchange diagram, and the exchange of the $\mathcal{O}(v^2)$ part of the propagator.  This last diagram may be equivalently described as the insertion of the $00$ part of the kinetic term.  The first approach is simpler calculationally if one knows the full time dependence of the propagator, which we do not in the current case.  However, even though we have not solved the time dependent propagator here, the diagram in question turns out to be proportional to the analogous diagram in GR, for which we know the full time-dependent propagator.  Thus, the relevant contribution is

\begin{equation}
S_\text{eff}^{(v^2)} = -\frac{m_1 m_2}{4\pi \alpha_2 \mpl^2} \int dt \frac{v_1^2 + v_2^2 }{R_{12}}
-\frac{m_1 m_2}{8\pi \alpha_2^2 \mpl^2} \int dt \left( \frac{\vec{v_1} \cdot \vec{v_2}}{R_{12}} + \frac{1}{R_{12}^3} (\vec{R}_{12} \cdot \vec{v}_1) (\vec{R}_{12} \cdot \vec{v}_2) \right) \, .
\end{equation}

%~~~~~~~~~~~~~~~~~~~~~~~~~~~~~~~~~~~~~~~~~~~~~~~~~~~~~~~~~~~~~~~~~~~~~~~~~~~~~~~~~~~~~~~~~~~~~~~~~~~~~~~~~~~~
\subsection{Breakdown of the Perturbative Expansion \label{ss:breakdown}}

As mentioned in Section~\ref{ss:outsiderv}, the effective action expansion we have outlined proves to be valid when one does not take into account distance hierarchies (e.g.\ the fact that the distance from the Earth to the Sun is much larger than the distance from the Earth to the Moon).  Let us now reconsider the expansion, taking this into consideration.  We will keep track of two separate distance scales: $r$, the distance of a dynamical object from the large central mass, and $R$, the separation between two arbitrary dynamical objects. 

Well within the Vainshtein radius, the propagator in \eqref{eq:phiphi_prop_full} now scales as
\begin{equation}
\langle \phi\phi \rangle \sim \frac{v}{r_v^{3/2} \sqrt{r}} \left( 1 + \frac{r}{R} \right) \, ,
\end{equation}
where the first term originates from the solution sourced by the $\delta$ function and the second term from the homogeneous solutions.  Thus, we must correct each appearance of $\phi$ by the addition of a factor $\sim \left(1+\frac{r}{R}\right)$.
Note that this is a conservative estimate in the situation where $R \ll r$, since we have treated time and space derivatives as scaling like $\frac{v}{r}$ and $\frac{1}{r}$, respectively.  When acting on the propagator given in~(\ref{eq:phiphi_prop_full}), derivatives will also produce terms which scale as $\frac{1}{R}$.  Such terms would introduce even larger corrections to instances of $\phi$ with derivatives acting on them.

\begin{figure}[ht]
\includegraphics[width=5in]{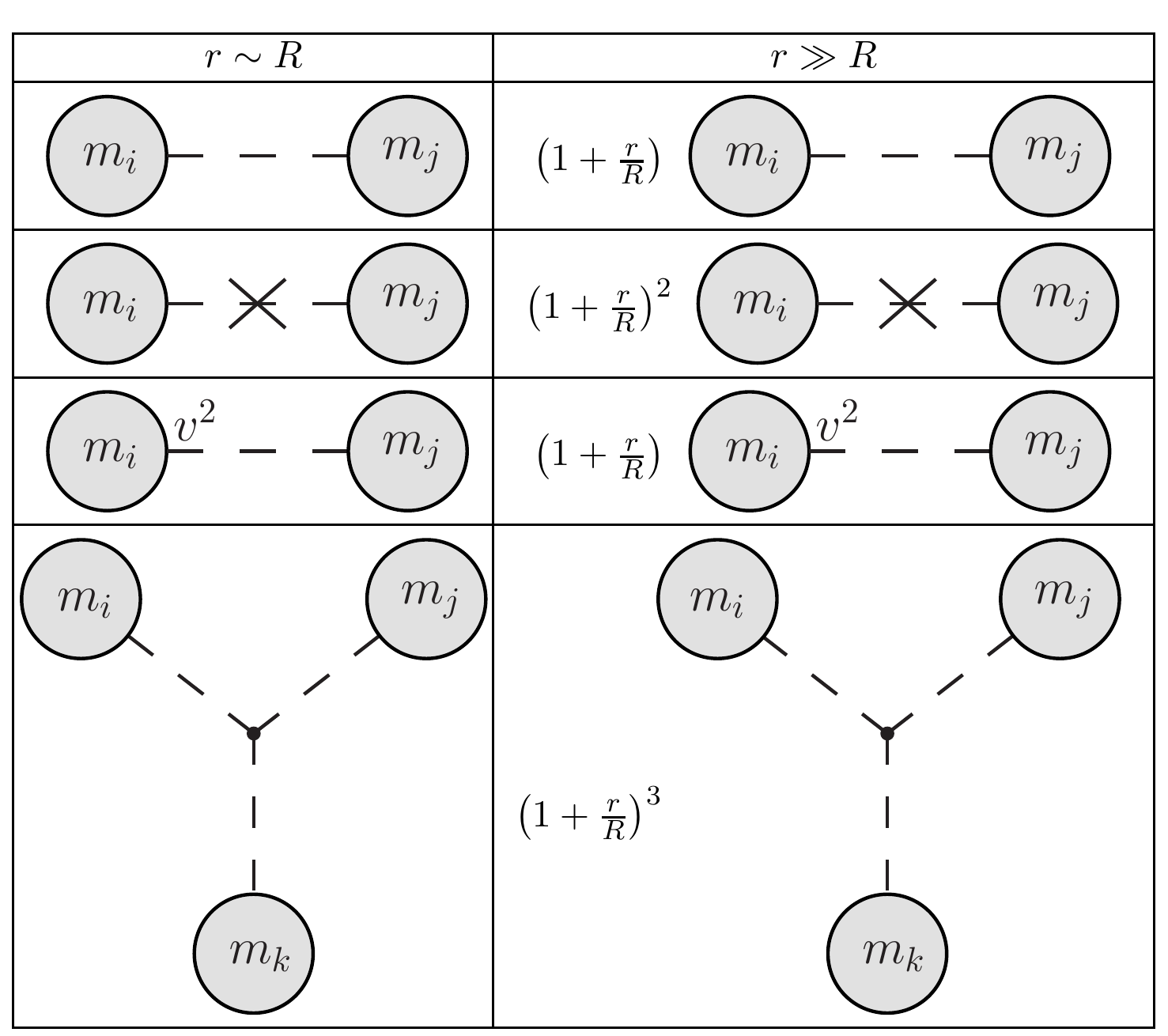}
\caption{\label{fig:distancehierarchies} A conservative estimate of corrections introduced by the consideration of object separations much smaller than the distance from either object to the large central mass.}
\end{figure}

These considerations correct the zeroth- and first-order (in either small parameter) diagrams as in Fig.~\ref{fig:distancehierarchies}  .
The new effective action amplitude is therefore $\hat{S}_0 = \widetilde{S}_0 \left(1+\frac{r}{R}\right)$ and the new expansion parameters are
\begin{equation}
v^2 \left(1+\frac{r}{R}\right) \hspace{2cm} 
\text{and} \hspace{2cm}
\hat{\epsilon} = \epsilon \left(1+\frac{r}{R}\right)^2 \, .
\end{equation}

This effect produces an enhancement in the overall strength of galileon interactions as seen in $\hat{S}_0$, as well as a breakdown of the expansion for sufficiently small separations between astrophysical objects.  As an example, consider the Earth-Moon system: the Sun is about $1000$ times more distant from the Earth than the Moon, and the Earth is about $10^6$ times less massive than the Sun.  Thus for the dynamics of the Earth-Moon system, the ``small" expansion parameter 
\begin{equation}
\hat{\epsilon} = \frac{m_\text{Earth}}{M_\text{Sun}} \left( 1 + \frac{r_\text{Sun-Earth}}{R_\text{Earth-Moon}} \right)^2
\sim \mathcal{O}(1) \, .
\end{equation}

In particular, this means that this formalism - approximating the galileon force by the first-order force obtained from the one-galileon-exchange diagram - cannot be used to constrain $\Lambda$ using lunar laser ranging experimental data, since the corrections to this force are large.  In the case of the force acting between the Earth and the Moon, diagrams with no velocity corrections but any number of galileon vertices all enter at the same order and hence any truncation of the expansion at finite order yields an error in the force of order the result calculated.

This conclusion is qualitatively similar to that reached in~\cite{piLagrangian}, which led to an even larger region in which the perturbative expansion is not valid.  In that view, the Earth's Vainshtein radius renormalized by the presence of the Sun is $\tilde{r}_{v,\oplus}^3 = (1\ AU)^3 \frac{M_\oplus}{M_\odot}$. Thus the quantity $\frac{m}{M} \left(\frac{r}{R}\right)^3 \sim \left(\frac{\tilde{r}_v}{R}\right)^3$ and the perturbative expansion about the background sourced by the Sun breaks down at a distance from the Earth of order the Earth's Vainshtein radius in the presence of the Sun.

Note that there is no analogous breakdown of the PPN expansion in GR, since $\left(\frac{v_\text{Earth}}{c}\right)^2 \sim 10^{-8}$ is smaller than the mass ratio.  Nor is there a problem for a system similar to the Earth and the Moon located outside the Sun's Vainshtein radius, as $\epsilon$ in this region has an additional small multiplicative factor $\left(\frac{r_v}{r}\right)^3$.

The importance of galileon self-interactions in determining astrophysical dynamics may mean that finite size effects, such as tidal forces acting on planetary bodies or the influence of their intrinsic multipole moments, could play a more significant role than their gravitational counterparts, relative to the lowest order forces between structureless test masses. We may take account of such effects through the addition of non-minimal terms to the world line action $\int dt (m/\mpl) \pi$. Of course, whenever we encounter an ultraviolet (UV) divergence when computing one of the relevant diagrams, the introduction of such counterterms is unavoidable. For comparison, in the case of GR, tidal effects first appear at $\mathcal{O}(v^{10})$ and are highly subdominant \cite{EFT}.

%~~~~~~~~~~~~~~~~~~~~~~~~~~~~~~~~~~~~~~~~~~~~~~~~~~~~~~~~~~~~~~~~~~~~~~~~~~~~~~~~~~~~~~~~~~~~~~~~~~~~~~~~~~~~
%~~~~~~~~~~~~~~~~~~~~~~~~~~~~~~~~~~~~~~~~~~~~~~~~~~~~~~~~~~~~~~~~~~~~~~~~~~~~~~~~~~~~~~~~~~~~~~~~~~~~~~~~~~~~
\section{Discussion\label{s:discussion}}
In this paper we have studied a number of topics crucial to a complete understanding of the effects of galileons on the dynamics of the solar system. We have obtained for the first time the first-order effects of the full quintic galileon theory. This includes the contribution to the precession of planetary perihelion due to the galileon field $\bar{\pi}$ of the Sun, as well as the static propagator for galileon fluctuations about $\bar{\pi}$. The inclusion of the higher interaction terms leads to a qualitatively different force law which yields a parametrically smaller perihelion precession than the cubic case. (The cubic case could potentially be observable with next-generation observations \cite{tests3}.) However, the presence of higher interaction terms exacerbates the superluminal propagation of radial perturbations as well as the very subluminal propagation of angular perturbations deep within the Vainshtein radius of the Sun.

To understand the effects of the cubic galileon theory on solar system dynamics, we have constructed a perturbative framework to calculate its effective action in the non-relativistic limit, about the background $\bar{\pi}$ sourced by the Sun. Apart from the typical speed $v$, there is an additional expansion parameter $\epsilon$ introduced by nonlinear galileon interactions. As a concrete example of the framework, we have calculated the first-order corrections in $v^2$ and in $\epsilon$ for the case where the objects whose galileon force we are interested in are outside the Vainshtein radius of the large central mass. Unfortunately, for the Earth-Moon system, we have shown that the additional expansion parameter $\epsilon$ becomes $\mathcal{O}(1)$, and thus that nonlinearities render the perturbative framework inadequate.

Even for well separated masses, where the perturbative expansion is valid, a concrete calculation of the corrections due to the nonlinear galileon interactions presents a technical challenge. If this could be done, it would yield a quantitative answer to the question, what is the Vainshtein radius of the Earth in the presence of the Sun? The closed-form propagator obtained in \eqref{eq:phiphi_prop_full} is only valid in the region deep within the Vainshtein radius; the full propagator in \eqref{GreensFunctionResult_Cubic} involves a spherical harmonic expansion. Corrections at $\mathcal{O}(\epsilon)$ involve an integral over {\it all space} of the propagator and hence require the full infinite sum form of the propagator. The integrals necessary to perform this calculation are not known, nor is there reason to believe it would yield a result that can be then summed into a closed-form solution.

\begin{acknowledgments}
We thank Kurt Hinterbichler for extensive discussions, and for collaboration in the early stages of this work. We also thank Lasha Berezhiani, Umberto Cannella, Justin Khoury, Denis Klevers, Tony Padilla and Rafael Porto for helpful conversations. This work was supported in part by the US Department of Energy and NASA ATP grant NNX11AI95G.
\end{acknowledgments}

%~~~~~~~~~~~~~~~~~~~~~~~~~~~~~~~~~~~~~~~~~~~~~~~~~~~~~~~~~~~~~~~~~~~~~~~~~~~~~~~~~~~~~~~~~~~~~~~~~~~~~~~~~~~~
%~~~~~~~~~~~~~~~~~~~~~~~~~~~~~~~~~~~~~~~~~~~~~~~~~~~~~~~~~~~~~~~~~~~~~~~~~~~~~~~~~~~~~~~~~~~~~~~~~~~~~~~~~~~~

\end{document}